\documentclass[twocolumn, twocolappendix]{aastex631}

\graphicspath{{./}{figures/}}

\newcommand{\mjybeam}{$\mathrm{mJy\,beam^{-1}}$}
\newcommand{\kms}{$\mathrm{km\,s^{-1}}$}
\newcommand{\Sersic}{S\'{e}rsic}
\newcommand{\barolo}{{$^\mathrm{3D}$\textnormal{Barolo}}}

\newcommand{\aco}{{$\alpha_\mathrm{CO}$}}
\newcommand{\Hmol}{{$\mathrm{H_2}$}}
\newcommand{\acounit}{{$M_\odot \mathrm{\,\left(K\,km\,s^{-1}\,pc^2\right)^{-1}}$}}

\usepackage{subfigure}
\usepackage{amsmath}
\usepackage{booktabs}
\usepackage{multirow}
\usepackage{ragged2e}

\begin{document}
\title{Dynamics of Molecular Gas in the Central Region of the Quasar I\,Zwicky\,1}
\author[0000-0001-7232-5355]{Qinyue Fei}
\affil{Kavli Institute for Astronomy and Astrophysics, Peking University, Beijing 100871, China}
\affiliation{Department of Astronomy, School of Physics, Peking University, Beijing 100871, China}

\author[0000-0003-4956-5742]{Ran Wang}
\correspondingauthor{Ran Wang}
\email{rwangkiaa@pku.edu.cn}
\affil{Kavli Institute for Astronomy and Astrophysics, Peking University, Beijing 100871, China}
\affiliation{Department of Astronomy, School of Physics, Peking University, Beijing 100871, China}

\author[0000-0002-8136-8127]{Juan Molina}
\affil{Department of Space, Earth and Environment, Chalmers University of Technology, Onsala Space Observatory, 439 92 Onsala, Sweden}

\author[0000-0002-4569-9009]{Jinyi Shangguan}
\affil{Max-Planck-Institut f\"{u}r Extraterrestrische Physik (MPE), Giessenbachstr., D-85748 Garching, Germany}

\author[0000-0001-6947-5846]{Luis C. Ho}
\affil{Kavli Institute for Astronomy and Astrophysics, Peking University, Beijing 100871, China}
\affiliation{Department of Astronomy, School of Physics, Peking University, Beijing 100871, China}

\author[0000-0002-8686-8737]{Franz E. Bauer}
\affil{Instituto de Astrof{\'{\i}}sica and Centro de Astroingenier{\'{\i}}a, Facultad de F{\'{i}}sica, Pontificia Universidad Cat{\'{o}}lica de Chile, Casilla 306, Santiago 22, Chile}
\affiliation{Millennium Institute of Astrophysics (MAS), Nuncio Monse{\~{n}}or S{\'{o}}tero Sanz 100, Providencia, Santiago, Chile} \affiliation{Space Science Institute, 4750 Walnut Street, Suite 205, Boulder, Colorado 80301}

\author[0000-0001-7568-6412]{Ezequiel Treister}
\affil{Instituto de Astrof{\'{\i}}sica and Centro de Astroingenier{\'{\i}}a, Facultad de F{\'{i}}sica, Pontificia Universidad Cat{\'{o}}lica de Chile, Casilla 306, Santiago 22, Chile}

\begin{abstract}
    We present a study of the molecular gas distribution and kinematics in the cicumnuclear region (radii $\lesssim 2\,$kpc) of the $z\approx0.061$ quasar I\,Zwicky\,1 using a collection of available Atacama Large Millimeter/submillimeter Array (ALMA) observations of the carbon monoxide (CO) emission. With an angular resolution of $\sim0.36''$ (corresponding to $\sim\,$400\,pc), the host galaxy sub-structures including the nuclear molecular gas disk, spiral arms, and a compact bar-like component are resolved. We analyzed the gas kinematics based on the CO image cube and obtained the rotation curve and radial distribution of velocity dispersion. The velocity dispersion is about 30\,\kms\ in the outer CO disk region and rises up to $\gtrsim 100\,$\kms at radius $\lesssim 1\,$kpc, suggesting that the central region of disk is dynamically hot. We constrain the CO-to-\Hmol\ conversion factor, \aco, by modeling the cold gas disk dynamics. We find that, with prior knowledge about the stellar and dark matter components, the \aco\ value in the circumnuclear region of this quasar host galaxy is $1.55_{-0.49}^{+0.47}$\,\acounit, which is between the value reported in ultra-luminous infrared galaxies and in the Milky-Way. The central 1\,kpc region of this quasar host galaxy has significant star formation activity, which can be identified as a nuclear starburst. We further investigate the high velocity dispersion in the central region. We find that the ISM turbulent pressure derived from the gas velocity dispersion is in equilibrium with the weight of the ISM.  This argues against extra power from AGN feedback that significantly affects the kinematics of the cold molecular gas.
\end{abstract}

\section{Introduction}

The scaling relationships between the supermassive black holes (SMBHs) and their host galaxies suggest that their early evolutionary progress are tightly coupled (e.g., \citealt{1998AJ....115.2285M,2000ApJ...539L...9F,2000ApJ...539L..13G,2013ARA&A..51..511K}). The active galactic nuclei (AGNs) represent the most active phase of the SMBH-galaxy co-evolution (\citealt{2007MNRAS.382.1415S, 2010MNRAS.402.1516K, 2010A&A...518L.155F, 2011ApJ...729L..27R, 2012ARA&A..50..455F, 2014A&A...562A..21C, 2017A&A...601A.143F, 2019MNRAS.483.4586F}). Cold molecular gas provides fuel for both star formation and SMBH growth (\citealt{2013ARA&A..51..105C, 2014MNRAS.441.1059V}). Studying the distribution and kinematics of the molecular gas is therefore crucial for understanding the physical process involved in the coevolution between SMBH and their host galaxies (\citealt{1991ApJ...370..158S, 2010A&A...518L.155F, 2011ApJ...733L..16S}). 

The low-order rotational transitions of carbon monoxide (CO) are the most common tracer for studies (e.g., \citealt{1989ApJ...337L..69B,2013ARA&A..51..105C,2017ApJ...846..159B,2018ApJ...859..144A,2019ApJ...887...24T,2021ApJ...908..231M,2021PASJ...73..257Y}). Massive molecular outflows reported in previous CO observations of AGN host galaxies are considered as evidence of negative AGN feedback, which expels gas and dust from the host galaxy (\citealt{2009ApJ...692.1623H, 2010A&A...518L.155F, 2014A&A...562A..21C, 2015A&A...580A...1M}). However, recent studies with large samples of optically selected quasars suggest that their host galaxies are falling on, and even above the main sequence of star-forming galaxies, with the host galaxy star formate rate (SFR) and SMBH accretion rate being tightly correlated (e.g., \citealt{2012ApJ...753L..30M,2013ApJ...773....3C,2017A&A...602A.123L,2021ApJ...906...38Z}). From an observational point of view, the impact of AGN feedback on host galaxy evolution is still under debate, and it is necessary to study the physical processes embedded in AGN host galaxies that govern the coevolution between SMBHs and host galaxies. 

Quasars, as the most luminous population of AGNs, are ideal targets for studying the impact of AGN feedback. \cite{2020ApJS..247...15S} observed the CO\,(2--1) line emission from a sample of 23 $z<0.1$ Palomar-Green \citep{1983ApJ...269..352S} quasars using the Atacama Compact (Morita) Array (ACA). \cite{2021ApJ...908..231M} provided follow-up ALMA observations of six PG quasars at $\sim$\,kpc-scale resolution to study the distribution and kinematics of molecular gas in their host galaxies. Their results suggest that quasar hosts and inactive star-forming galaxies have similar gas fractions \citep{2020ApJS..247...15S}, but more centrally concentrated \citep{2021ApJ...908..231M}; luminous quasars do not efficiently remove cold gas from the host galaxy. 

Accurate measurements of the cold gas mass are key to understand the ISM evolution in quasar host galaxies. The molecular gas masses are mainly measured by using the line luminosities of the low-$J$ CO transitions based on assumptions of the CO (1--0) luminosity-to-mass conversion factor \aco, i.e., $M_\mathrm{H_2}=\alpha_\mathrm{CO}L'_\mathrm{CO\,(1-0)}$. A Millky-Way-like value of 3.1 \citep{2013ApJ...777....5S} is usually assumed for local Seyferts and quasars that are hosted in spiral galaxies \citep{2006AJ....132.2398E,2020ApJS..247...15S,2021ApJS..252...29K}, while the ULIRG-like value of $\alpha_\mathrm{CO}=0.8\,$\acounit \citep{1998ApJ...507..615D} is also considered for AGNs that are hosted in starburst systems \citep{2012ApJ...750...92X}. Studies of star-forming galaxies from local to high-$z$ suggest that the \aco\ factor varies over a wide range (\citealt{1987ApJ...319..730S,2006A&A...454..781L,2011MNRAS.418..664N,2012MNRAS.426.2601P,2013ApJ...777....5S}) and depends on the metallicity of the ISM \citep{1997A&A...328..471I,2010ApJ...716.1191W,2011ApJ...737...12L}. However, there are still few direct measurements of \aco\ in quasar host galaxies \citep{2020ApJS..247...15S}. 

High resolution molecular CO line imaging with ALMA opens a unique opportunity to measure the gas dynamics in the nearby quasar host galaxies (e.g., \citealt{2019ApJ...887...24T}). The rotation curve traced by the CO line velocity field constrains the dynamical mass of the host galaxy, allowing a detailed study of the mass budget from the gas and stellar content and providing an independent way to measure the \aco\ factor. In this work, we present a case study of the quasar I Zwicky 1 (hereafter I\,Zw\,1). The CO\,(2--1) emission from its host galaxy was observed by ALMA at $\sim 400$\,pc scale, the highest spatial resolution (by a factor of $\sim2-3$) among the six objects presented in \citet{2021ApJ...908..231M}, which allows us to resolve the gas content in the central few kpc region. The high-resolution data allows the possibility of dynamical analysis, which is widely used in investigating the accurate mass-to-light ratio in galaxies (e.g., \citealt{2008AJ....136.2648D}).

I\,Zw\,1 possesses one of the most complete sets of multi-wavelength spectral energy distribution (SED) data coverage (\citealt{1976ApJ...208...37P, 1989ApJS...70..257B, 2004A&A...417...29G, 2008ApJ...675...83B, 2018MNRAS.480.2334S, 2019ApJ...886...33L}). Spectroscopic observations indicate that it is a narrow line Seyfert 1 system with $\mathrm{FWHM_{H\beta}}=1400\,$\kms \citep{1977ApJ...215..733O}, and a BH mass of $9.30_{-1.38}^{+1.26}\times 10^6M_\odot$ given by reverberation mapping \citep{2019ApJ...876..102H}. With a bolometric luminosity of $L_\mathrm{bol}=3\times 10^{45}\,\mathrm{erg\,s^{-1}}$, I\,Zw\,1 is quantified as a super-Eddington source with $\lambda_\mathrm{Edd}=2.58$. Long term X-ray monitering indicates the existence of an ultra fast outflow in the nucleus of I\,Zw\,1 \citep{2022ApJ...931...77D}. Detailed morphological analysis based on Hubble space telescope (HST) $0.'' 1$ resolution image showed a prominent pseudo-bulge (\Sersic\ index $n\approx 1.69$, effective radius $r_e\approx1.6\,$kpc), and a relatively faint and extended disk \citep{2021ApJ...911...94Z}. The pseudo-bulge implies a black hole to bulge mass of $\sim10^{-4}$, smaller than that of classical bulges and elliptical galaxies by a factor of 50 \citep{2019ApJ...876..102H}. The SED decomposition analysis yields a far-infrared (FIR) luminosity of $\log L_\mathrm{FIR}/L_\odot=11.94\pm0.30$ \citep{2018ApJ...854..158S}, in the range of Luminous Infrared Galaxies (LIRGs). The star formation activity distribution was confirmed with newly developed integrated field units (IFU) observations (\citealt{2021A&A...646A.101P, 2022ApJ...935...72M, 2022arXiv220903380L}). The significant star formation activity is also confirmed by the combination of optical and sub-mm observations \citep{2022arXiv221205295M}. Previous IRAM and ALMA observations already suggested that I\,Zw\,1 has a rich molecular gas reservoir mainly concentrated in its circumnuclear zone (\citealt{1989ApJ...337L..69B,1994ApJ...424..627E, 1998ApJ...500..147S, 2019ApJ...887...24T}).

This paper is organized as follows: In Section \ref{sec2} we present the available ALMA archival data of CO\,(1--0) and CO\,(2--1) observations and describe the data reduction. In section \ref{sec3} we model the molecular gas distribution and kinematics. In Section \ref{sec4}, we model the gas dynamics and estimate the mass of each component, with prior knowledge of stellar distribution and dark matter halo properties. In section \ref{sec5} we discuss the CO emission line ratios and the surface density of SFRs, and investigate whether we detect significant AGN feedback. We summarize in Section \ref{sec6}. For standard cosmological parameters of $\Omega_m=0.308$, $\Omega_\Lambda=0.692$, and $H_0=67.8\,\mathrm{km\,s^{-1}\,Mpc^{-1}}$ \citep{2016A&A...594A..13P}, the redshift of $z=0.06115$ corresponds to a luminosity distance of 283\,Mpc.

\section{CO data of I\,Zw\,1}
\label{sec2}

\begin{table*}
    \centering
    \caption{ALMA CO\,(2--1) observations of I\,Zw\,1}
    \begin{tabular}{ccccccc}
        \hline
        \hline
        Project & Date of & Observational & \multirow{2}{*}{Configuration} & Antenna & On-target & \multirow{2}{*}{References} \\
        Code & Observation & band & & number & time & \\
         & & & & & (s) \\
        (1) & (2) & (3) & (4) & (5) & (6) & (7)\\
        \hline
        2017.1.00297.S & Nov. 2017 & Band 6 & ACA & 11 & 8991 & \cite{2020ApJS..247...15S}\\
        2018.1.00006.S & Nov. 2018 & Band 6 & C43--5 & 44 &699 & \cite{2021ApJ...908..231M}\\
        2018.1.00699.S & Oct. 2018 & Band 6 & C43--5 & 45 & 2992 & \cite{2022arXiv220903380L}\\
        \hline
    \end{tabular}
    \\
    \vspace{1mm}
    \justify{\justify \textsc{Note} --- (1) The project code of ALMA observations. (2) The date of ALMA observations. (3) The ALMA band used during the observation. (4) The configuration of ALMA during the observation. (5) The number of antennas that are used during the observation. (6) The total on-target time of observation. (7) Papers that first report the observation.}
    \label{table1}
\end{table*}

We collect available observations of the CO\,(2--1) line emission of I\,Zw\,1 from the ALMA archive. The final data are combined from three ALMA programs, 2017.1.00297.S, 2018.1.00006.S (PI: Franz Bauer) and 2018.1.00699.S (PI: Pereira Santaella, Miguel) (\citealt{2020ApJS..247...15S, 2021ApJ...908..231M, 2022arXiv220903380L}). The first observation is our Atacama Compact (Morita) Array (ACA) survey, with 2.5 hours on-source integration time and an angular resolution of 7$''$ \citep{2020ApJS..247...15S}. The second observation is our follow-up high-angular resolution observation, with about 11 minutes on-source time and an angular resolution of 0.4$''$ \citep{2021ApJ...908..231M}. The third observation is a part of the ``Physics of ULIRGs with MUSE and ALMA'' (PUMA; \citealt{2022arXiv220903380L}) project, with 50 minutes on-source time and an angular resolution of 0.3$''$. We list the details of these observations in Table \ref{table1}.

We use the Common Astronomy Software Application (CASA) version 5.6.1 \citep{2007ASPC..376..127M} to reduce the ALMA observation data. All of these observations are concatenated with the CASA task \texttt{concat}. The continuum data are fitted and subtracted with CASA task \texttt{uvcontsub}. We then imaged and cleaned the line data cube and continuum data with Briggs weighting (robust $=0.5$) and a stop threshold 2.5 times the root mean square (rms) of the off-source channels. For CO\,(2--1) emission line we set a channel resolution of 7.812\,MHz, which corresponds to $\sim$11\,\kms\ at $z=0.061$. We set gridder$\,=\,$mosaic during \texttt{tclean}, and employ the \texttt{auto-multithresh} masking procedure \citep{2007ASPC..376..127M}. We set the noise-, sidelobe-, and lownoise-threshold as 4.25, 2.0, and 1.5 as recommended by CASA guideline.\footnote{\url{https://casaguides.nrao.edu/index.php/Automasking_Guide}} The other additional parameters were not modified. Finally, we obtain a CO\,(2--1) datacube with a synthesized beam size of $0.36''\times0.32''$, and typical channel root mean square (rms) noise of 0.28$\,\mathrm{mJy\,beam^{-1}}$. We derive the velocity-integrated flux map, intensity-weighted velocity, and velocity dispersion maps using the CASA task \texttt{immoments}. The beam size of the 1.3\,mm continuum is $0.31''\times0.28''$, and the rms of the continuum map is 0.012$\,$\mjybeam.

We also investigate the surface density distribution of CO\,(1--0), in order to image the CO\,(2--1)/CO\,(1--0) emission line ratio of this target. We build the ALMA CO\,(1--0) data, which is adapted from the ALMA program 2015.1.01147.S \citep{2019ApJ...887...24T}. We reduce the data following the procedure described previously. The beam size of CO\,(1--0) data is $0.62''\times0.57''$.

\section{Results and Analysis}
\label{sec3}

\subsection{Distribution of the molecular gas}
\label{3.1}

\begin{table*}
    \centering
    \def\arraystretch{1.5}
	\setlength\tabcolsep{4pt}
    \caption{Fitting parameters of CO\,(2--1) intensity map}
    \begin{tabular}{cccccc}
        \hline
        \hline
         & $I_e$ & $R_e$ & $n$ & $b/a$ & $\phi_\mathrm{s}$\\
         & ($\mathrm{Jy\,beam^{-1}\,km\,s^{-1}}$) &  ($''$) & & & ($^\circ$)\\
         & (1) & (2) & (3) & (4) & (5) \\
        \hline
        CND & $0.44^{+0.01}_{-0.01}$ & $1.30^{+0.01}_{-0.01}$ & $0.48^{+0.01}_{-0.01}$ & $0.80_{-0.01}^{+0.01}$ & $141.90_{-0.40}^{+0.40}$ \\
        Bar & $2.37_{-0.05}^{+0.05}$ & $0.50_{-0.01}^{+0.01}$ & $0.30_{-0.20}^{+0.20}$ & $0.32_{-0.01}^{+0.01}$ & $33.65_{-0.11}^{+0.11}$ \\
        \hline
        \hline
         & $I_G$ & $\mathrm{FWHM_{maj}}$ & $\mathrm{FWHM_{min}}$ & $\phi_\mathrm{G}$\\
         & ($\mathrm{Jy\,beam^{-1}\,km\,s^{-1}}$) & (mas) & (mas) & ($^\circ$)\\
          & (6) & (7) & (8) & (9) \\
        \hline
        Core & $29.90_{-0.70}^{+0.73}$ & $221.35_{-0.01}^{+0.01}$ & $122.45_{-0.01}^{+0.01}$ & $16.59_{-1.56}^{+1.36}$\\
        \hline
    \end{tabular}
    \\
    \vspace{1mm}
    \justify{\justify \textsc{Note} --- (1) Intensity at effective radius. (2) Effective radius. (3) \Sersic\ index. (4) The minor-to-major axis ratio. (5) Position angle of the major axis. $\mathrm{North}=0^\circ$, $\mathrm{East}=90^\circ$. (6) Amplitude of Gaussian function. (7) and (8) FWHM of the major and minor axis. (9) The position angle of the major axis of the Gaussian component.}
    \label{table2}
\end{table*}

\begin{figure*}
    \centering
    \includegraphics[width=\linewidth]{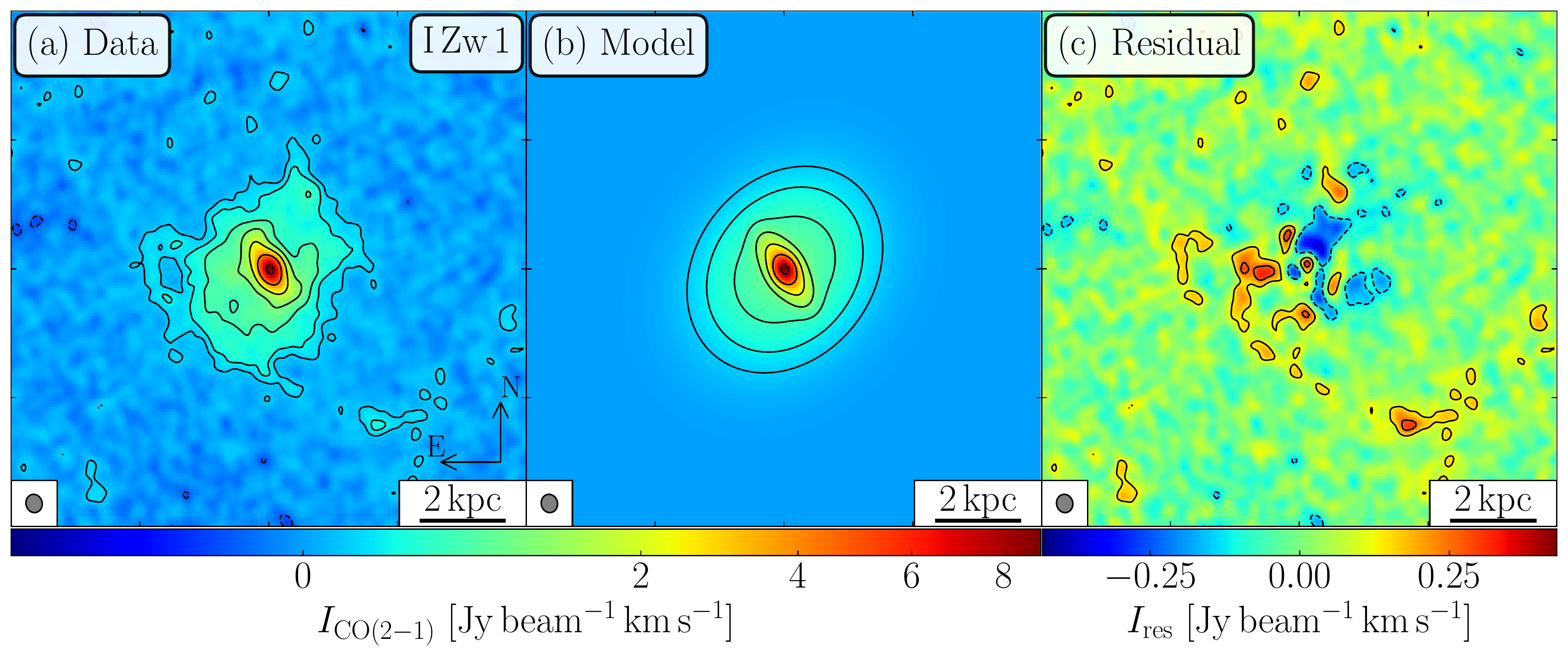}
    \caption{The comparison between observed and modeled intensity maps of the CO\,(2--1) line emission. Contour levels in each panel correspond to $[-1,1,2,4,8,16,32]\times 3\sigma$, where $\sigma$ is the rms, with the value of $0.043\,\mathrm{Jy\,beam^{-1}\,km\,s^{-1}}$. Panel (a) and (b) represent the velocity-integrated map of the data and the model, respectively. Panel (c) represents the residuals between the data and the model. The North and East direction is shown as arrow at the lower right corner in panel (a). The synthesized beam ($0.36''\times0.32''$) is plotted at the bottom left corner of each panel.}
    \label{intensity_fitting}
\end{figure*}

We present the velocity-integrated intensity map in Figure \ref{intensity_fitting} (a). The CO\,(2--1) line emission in I\,Zw\,1 traces a disk with a diameter of $\sim5\,$kpc, which is consistent with the source size of the CO\,(1--0) line emission \citep{2019ApJ...887...24T}. In Figure \ref{intensity_fitting} (a), we note that the central contours (above 24$\sigma$) are elongated along the northeast-southwest direction while the outer lower surface brightness region has a different major axis position angle. This indicates that the molecular gas disk can be described morphologically by two components, one extremely compact bar-like structure and an extended circumnuclear disk (CND), extends up to $\sim$1\,kpc at an position angle of $\sim30^\circ$. Such elongated structure could also be a massive bipolar gas outflow; however, the further kinematic analysis do not show evidence of any non-circular motions (Sec. \ref{3.2}). The large intensity gradient in the nucleus also implies that the CO emission may also exhibit a central compact core component unresolved by ALMA.

The complex molecular gas distribution described above can be well-described by fitting the CO\,(2--1) line intensity map with three components: two \Sersic\ \citep{1963BAAA....6...41S} components for the extended emission (equation \ref{sersic}) and one Gaussian component for the unresolved core (equation \ref{gaussian}):
\begin{align}
    I_s(r) &= I_e\exp\left\{-b_n\left[\left(\frac{r}{r_e}\right)^{1/n}-1\right]\right\},\label{sersic}\\
    I_g(r) &= I_G\exp\left\{-\frac{r^2}{2\sigma^2}\right\},
    \label{gaussian}
\end{align}
,where $I_e$ is the surface brightness measured at $r_e$, the effective radius, $n$ is the \Sersic\ index, and $b_n$ is the numerical coefficients that ensures $r_e$ corresponding to the half-light radius \citep{1963BAAA....6...41S}. We use these two \Sersic\ profiles to describe the bar-like structure and disk component, respectively, and use the Gaussian profile to describe the central compact core. We built this three-component model with Astropy \citep{2013A&A...558A..33A}, which is then convolved with the observation synthesized beam to produce the model of the observed line intensity map. The three-component model contains sixteen free parameters, including $I_e,\,r_e,\,n,$ minor-to-major axis ratio ($b/a$), position angle ($\phi_\mathrm{s}$) for each of the two \Sersic\ components, Gaussian amplitude ($I_G$), full-width at half-maximum along the major and minor axes ($\mathrm{FWHM_x},\,\mathrm{FWHM_y}$), position angle of the major axis ($\phi_\mathrm{G}$) for the Gaussian component, and the center location $(x_0,\,y_0)$ that is shared with all the three components.

To find the best-fitting model we use the Python package \texttt{emcee} \citep{2013PASP..125..306F}. The \texttt{emcee} package implements the affine-invariant ensemble sampler for Markov chain Monte Carlo (MCMC) sampling method to sample the posterior probability distribution function (PDF). We optimize the log-likelihood function:
\begin{align}
    \log \mathcal{L}\equiv -\frac{1}{2}\sum_i^N \left[\frac{(z_i-z_i^m)^2}{\sigma_i^2}+\ln(2\pi\sigma_i^2)\right],
\end{align}
where $z_i$ denotes the surface brightness at each pixel, $\sigma_i$ is the $1\sigma$ noise, and $z_i^m$ correspond to the model value at same pixel. The best fitting model along with residuals are shown in Figure \ref{intensity_fitting}, and the best fitting parameters are presented in Table \ref{table2}.

With the assumption that the CND and stellar disk are coplanar, we estimate the inclination angle of the disk following the formula in \cite{1926ApJ....64..321H},
\begin{align}
    \cos ^2 i = \frac{(b/a)^2-q_0^2}{1-q_0^2},
\end{align}
where $q_0$ is the intrinsic galaxy thickness, $b/a$ is the minor-to-major axis of the CND. We assume $q_0=0.14$ for the molecular gas disk, which is similar with that reported for edge-on galaxies at low redshifts \citep{2015MNRAS.451.2376M}. We obtain a host galaxy inclination $i=38\,^\circ$.

From Figure \ref{intensity_fitting}, we can see that our model presents a reasonable description of the gas distribution in the circumnuclear scale. The residuals are likely to be produced by partially resolved out spiral arms as can be seen at the South West edge.

\subsection{\barolo\ fitting}
\label{3.2}
\begin{figure*}
    \centering
    \includegraphics[width=\linewidth]{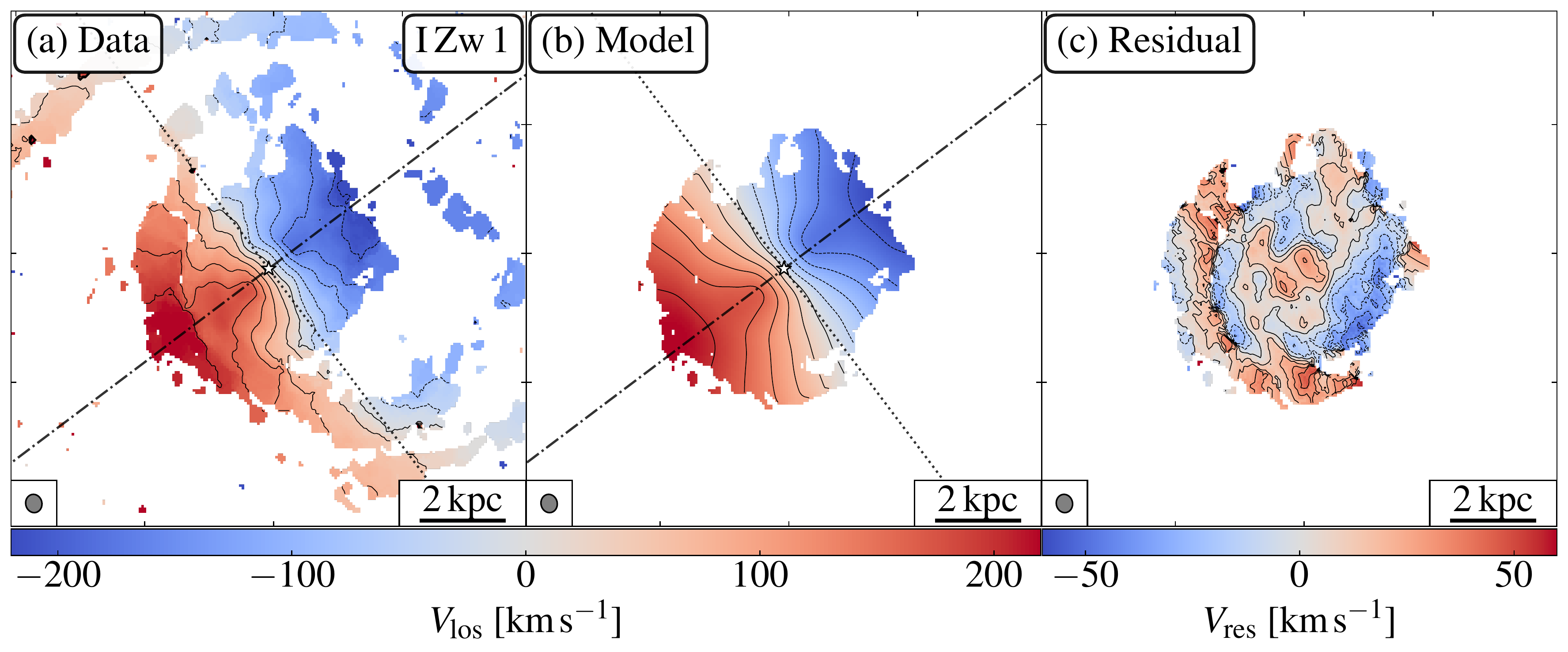}
    \includegraphics[width=\linewidth]{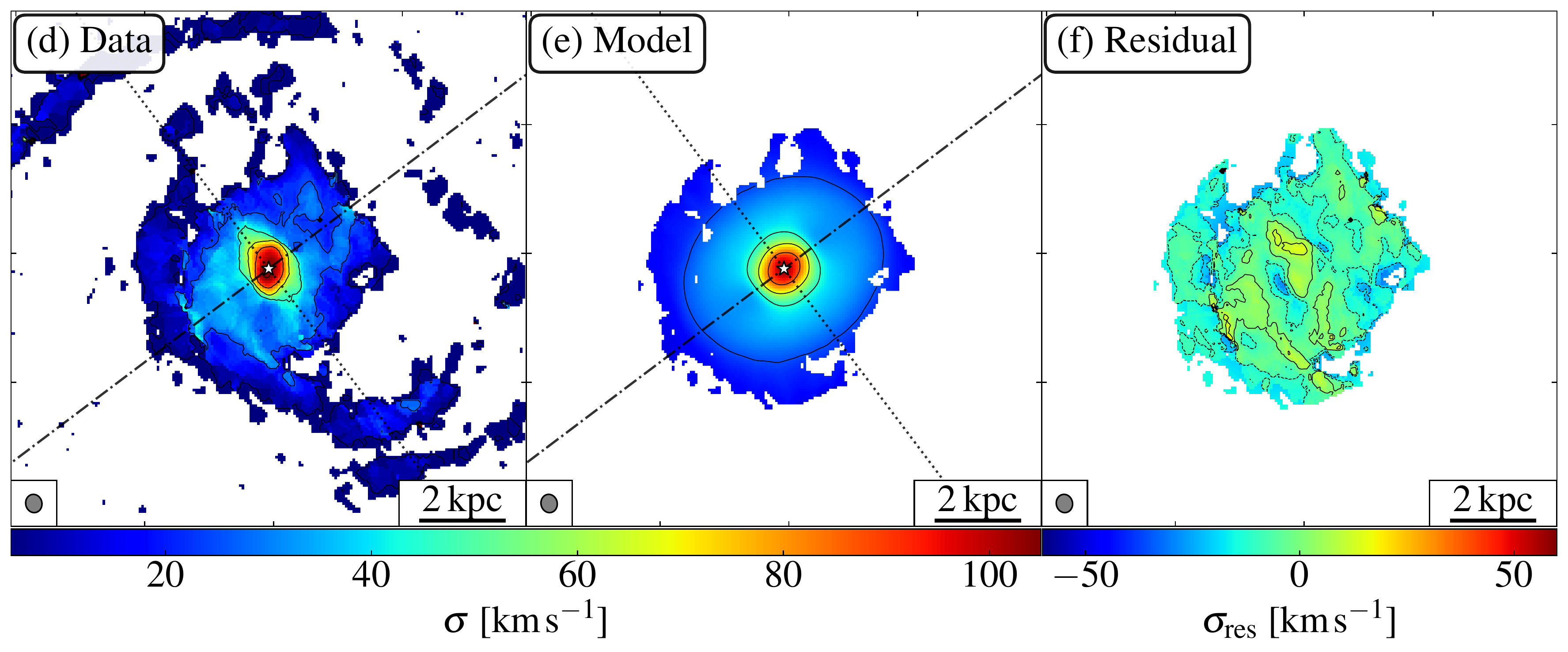}
    \caption{The results of \barolo\ fitting. Panel (a) and (b) represent the line-of-sight velocity map of the CO\,(2--1) data and the best-fitting model given by \barolo. Panel (c) represents the residual between the observation and the model. Panel (d) and (e) represent the velocity dispersion of data and \barolo\ model. Panel (f) represents the residual of the velocity dispersion map. Contours in panel (a) and (b) start from -200\,\kms\ and in step of 40\,\kms. Contours in panels (d) and (e) are from 5\,\kms\ and in step of 40\,\kms. Contours in residual maps range from -50\,\kms\ to 50\,\kms and in step of 20 \,\kms. The white star in each sub-panel indicates the kinematics center. The dash-dotted and dotted lines represent the major- and minor-kinematic axes.}
    \label{kinematic_fitting}
\end{figure*}

\begin{figure*}
    \centering
    \includegraphics[width=\linewidth]{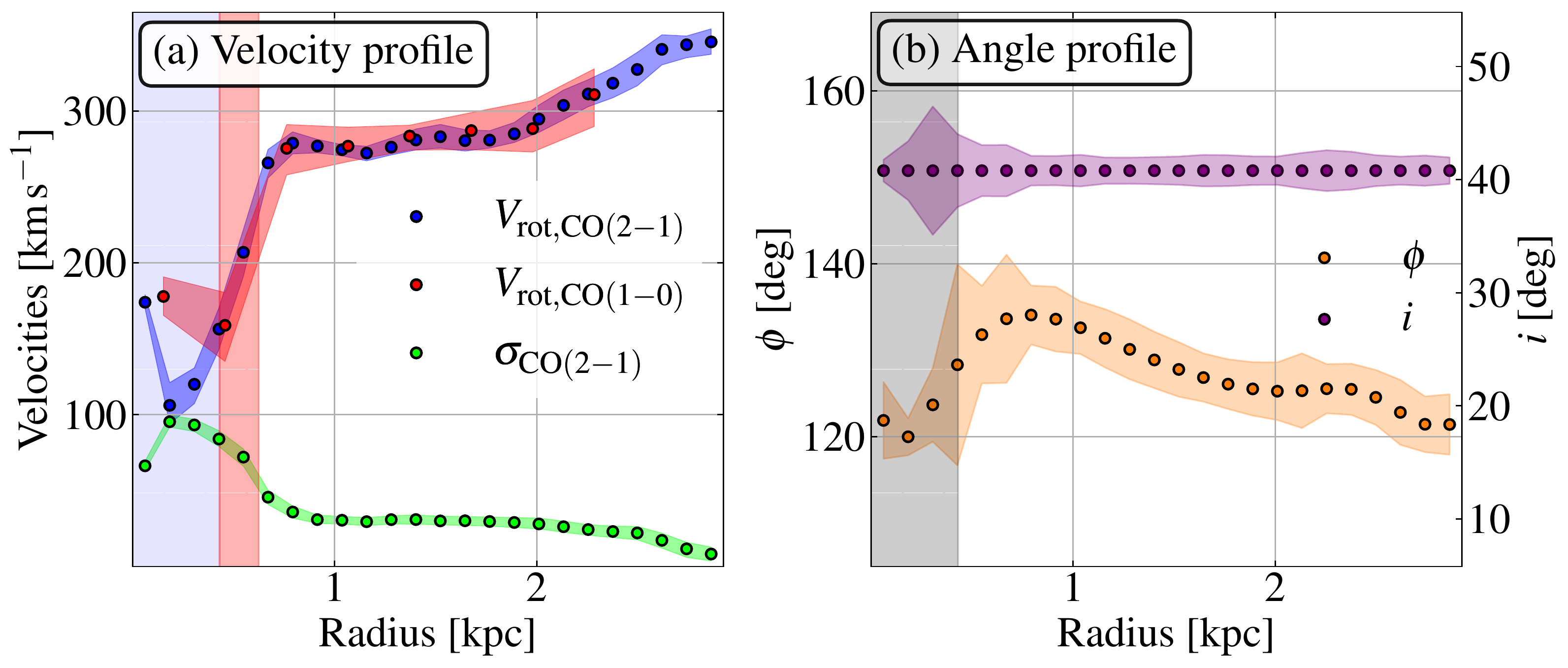}
    \caption{The rotation velocity, velocity dispersion, inclination angle, and kinematic position angle derived from \barolo\ fitting to the CO\,(2--1) data. Panel (a) represents the rotation velocities (blue points) and velocity dispersions (green points) as a function of radius. Red points represent the rotation velocities extracted from the CO (1-0) data at a resolution of $\sim600$\,pc for comparison. Panel (b) represents the position angle of the kinematic major axis (purple points) and the inclination angle (orange points) as a function of radius. The uncertainties of parameters are shown as colored shades. The blue and red vertical shaded region represents the beam size of the observation for CO(2--1) and CO(1--0) observations.}   
    \label{barolo_results}
\end{figure*}

The intensity-weighted velocity and velocity dispersion maps of CO\,(2--1) line emission are shown in panel (a) and panel (d) in Figure \ref{kinematic_fitting}. As discussed in \cite{2021ApJ...908..231M}, the gas velocity field is dominated by circular rotation. The velocity dispersion in the outer region is almost constant ($\sim30\,\mathrm{km\,s^{-1}}$) with small variations along the radius, while in the inner 1\,kpc region, the velocity dispersion rises up to 100\,\kms.

Assuming that the non-circular motions are negligible, we fit the velocity field with a tilted ring model \citep{1974ApJ...193..309R}. The rotating disk is decomposed into a series of thin rings, and the kinematic properties of each ring can be described by seven parameters:
\begin{itemize}
    \item [1.] $(x_0, y_0)$: the sky coordinates of the ring center;
    \item [2.] $V_\mathrm{sys}$: the systematic velocity of the center of the ring related to the observer;
    \item [3.] $V_\mathrm{rot}(R)$: the rotation velocity of the ring;
    \item [4.] $\sigma$: the velocity dispersion of the ring;
    \item [5.] $\phi(R)$: the position angle of the kinematic major axis on the receding half of the galaxy, with respect to the north direction;
    \item [6.] $i(R)$: the inclination angle between the normal to the ring and the line-of-sight, $\mathrm{Inc.}=0^\circ$ represents a face-on disk;
    \item [7.] $z_0$: the scale height of the gas layer.
\end{itemize}

The line-of-sight velocity field [$V_\mathrm{los}(x,y)$] that we observed is related to the above parameters:
\begin{align*}
    V_\mathrm{los}(x,y) = V_\mathrm{sys} + V_\mathrm{rot}(R) \sin i(R) \cos \theta\\
    \cos\theta = \frac{-(x-x_0)\sin\phi + (y-y_0)\cos\phi}{R},
\end{align*}
where $R$ is the radius of each ring.

In order to obtain the intrinsic kinematics of molecular gas in this galaxy, we model the ALMA datacube using the 3D-Based Analysis of Rotating Objects from Line Observations (\barolo, version 1.6; \citealt{2015MNRAS.451.3021D}). \barolo\ fits the three dimensions of data cubes with a tilted-ring model. By directly modeling the data cube instead of the 2D velocity map, it fully accounts for the beam smearing effect, providing a reasonable model of the intrinsic circular velocity and velocity dispersion field for circular rotating systems (see \citealt{2015MNRAS.451.3021D}, for more details).

We fit the gas kinematics with \barolo\ in two steps following the procedure described in \cite{2018ApJ...859..144A}, but with some revisions. In the first step, we set the galaxy center $(x_0, y_0)$, systematic velocity $V_\mathrm{sys}$, rotation velocity $V_\mathrm{rot}$, velocity dispersion $\sigma$, position angle $\phi$, inclination $i$, and disk height $z_0$ parameters to be free. We adopt a ring width of 0.1 arcsec in the fitting, roughly one-third of the beam size. Initial guesses for the position angle and inclination are adopted from the morphological model results ($\phi=142\,^{\circ}$ and $i=38\,^{\circ}$; Section \ref{3.1}). Initial guesses for the kinematic center are set to be the same as the morphological center. We find that the output kinematic centers of each ring from this initial center are almost constant along the CND, in radii between 0.8 and 2.1\,kpc. However, the fitting kinematic centers of the inner and outer rings show large scatter and uncertainties, which is possibly due to the limited resolution, poor sampling, and complex dynamics caused by compact central bar-like structure and possible companion interaction in the outer region \citep{2020ApJS..247...15S}. Other fitting parameters, such as $V_\mathrm{sys}$, are almost constant within the CND. During the second fitting step, we fix the kinematic center and systematic velocity to the mean values over CND scale that are obtained from the first fitting step, then to fit the rotation velocities, velocity dispersions, position angles, and inclination angles for each ring. The \barolo\ best-fitting results are shown in Figure \ref{barolo_results}.

The 3-D model successfully describes the CND cold gas kinematics, with rms model residuals $\approx 20$\,\kms\ and $\approx 10$\,\kms\ for the rotation velocity and velocity dispersion fields, respectively. The latter is comparable to the velocity resolution of the observation. 

\subsection{Global kinematics}
\label{3.3}
In panels (a) and (b) of Figure \ref{barolo_results} we show the velocity and angles radial profiles derived by \barolo. The rotation velocity rises to the flattened part at $\sim0.8$\,kpc, with a value of $\sim 270\,$\kms, and slightly increases toward a larger radius in the spiral arm region ($r>2.1\,$kpc). In this region, the velocity dispersion is $\sim30$\,\kms, which indicates a cold gas disk with $V/\sigma\approx 9$.

The velocity dispersion profile increases from 30$\,$\kms\ at $\sim0.8\,\mathrm{kpc}$ to 100$\,$\kms\ at $\sim0.3\,\mathrm{kpc}$. A similar high central velocity dispersion was also reported in \cite{2021ApJ...908..231M}. This enhanced velocity dispersion in the central region is unlikely spurious due to the beam smearing effect as \barolo\ is designed to take this into account \citep{2015MNRAS.451.3021D}. To further check this, we build a mock disk model adopting an intrinsic rotation curve of I\,Zw\,1 and the constant velocity dispersion of $\sigma=30\,$\kms\ at all radii. We set the inclination angle and position angle equal to 41$^\circ$ and 130$^\circ$, which are the same as those in I\,Zw\,1. We then simulate the ALMA observational data cube in CASA using the \texttt{simobserve} task and fit the mock data cube with \barolo. Through our simulated data, we find that the beam-smearing effect can only increase the velocity dispersion value by a factor of $\sim 1.3$, insufficient to account for the observational increase of a factor of $\sim 4$. Thus we conclude that the velocity dispersion is intrinsically high in the center region. (see detailed discussion in Appendix \ref{A}). We investigate the origin of such high-velocity dispersion in Section \ref{5.3}.

Naturally, \barolo\ poorly fits the datacube in the zones where the inner spiral arms are present, but those regions display significant non-circular motions reflecting the local perturbation of the gravitational potential field. Additional kinematic components may be included for a more accurate model for those regions. \cite{2021ApJ...908..231M} investigated the non-circular motion of the CO\,(2--1) line velocity field of this object with \texttt{KINEMETRY} \citep{2006MNRAS.366..787K}, finding that non-circular motions are negligible. However, the compact bar-like structure presented in our morphology analysis could still introduce non-circular components in the velocity field of the very central region ($\lesssim 1\,$kpc), which requires higher resolution observations to fully resolve the bar-like structure kinematics.

\subsection{Continuum}
\label{3.4}

\begin{figure}
    \centering
    \includegraphics[width=\linewidth]{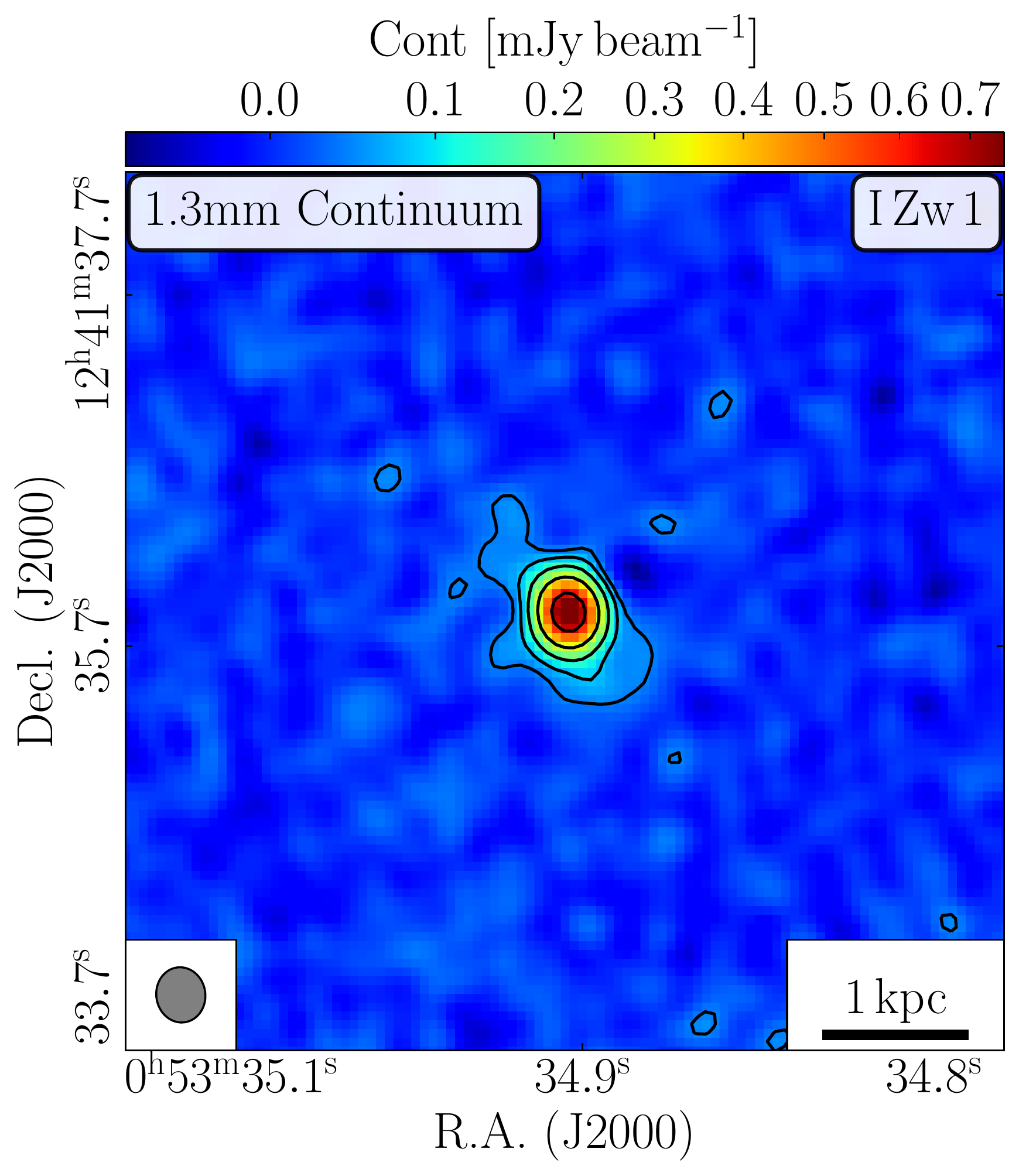}
    \caption{The 1.3\,mm continuum map of I\,Zw\,1 host galaxy. The contour levels correspond to $[-1,\,1,\,2,\,4,\,8,\,16,\,32] \times 3\,\sigma$, where $\sigma=0.012\mathrm{mJy\,beam^{-1}}$.  The synthesized beam ($0.31''\times0.28''$) is plotted at the lower left corner.}
    \label{continuum}
\end{figure}

Figure \ref{continuum} shows the 1.3\,mm map of this galaxy. Although the size of the continuum map is more compact than that of the CO\,(2--1) line-emitting region, we still can see the structure which is elongated in a northeast-to-southwest direction. The continuum source has a position angle similar to that of the molecular bar-like structure. The total flux density within the 3$\sigma$ ``contour'' region is $1.1\pm0.1\mathrm{mJy}$, which covers 73\% of the continuum flux density obtained from the previous ACA observation, and 34\% of the continuum flux density from global far-IR SED fitting prediction (\citealt{2018ApJ...854..158S,2020ApJS..247...15S}).

We fit the size of the continuum using the \texttt{imfit} task in CASA, which performs synthesized beam deconvolution and two-dimensional (2D) Gaussian fitting to the images. The resulting deconvolved full width at half maximum (FWHM) of major axis and minor axis sizes are $0.174\pm0.014\,''$ and $0.100\pm0.017\,''$, corresponding to $(0.21\pm0.02)\times(0.02\pm0.01)\,\mathrm{kpc}^2$, with a position angle of $23.5\pm9.9\,^\circ$.

\section{Dynamical modeling and the CO-to-H$_2$ conversion factor}
\label{sec4}

\begin{table*}
        \centering
        \caption{Constraints and results of dynamical parameters}
        \begin{tabular}{ccccccccc}
            \hline
            \hline
             & $\log M_\mathrm{b}$ & $r_\mathrm{e,b}$ & $n$ & $\log M_\mathrm{d}$ & $r_\mathrm{e,d}$ & $\log f_*$ & $\log c$\,$^a$ & $\alpha_\mathrm{CO}$\\
             & $M_\odot$ & kpc & & $M_\odot$ & kpc & & & $M_\odot\left(\mathrm{K\,km\,s^{-1}pc^2}\right)^{-1}$\\
             & (1) & (2) & (3) & (4) & (5) & (6) & (7) &(8) \\
            \hline
            \textbf{Prior} & (9.5, 12.5) & $1.62\pm0.05$ & $1.69\pm0.05$ & (9.1, 12.1) & $10.97\pm0.50$ & (-2.25, -1.30) & (0.52, 1.12) & (0, 20)\\
            \textbf{Posterior} & $10.70_{-0.10}^{+0.08}$ & $1.61_{-0.05}^{+0.05}$ & $1.70_{-0.05}^{+0.05}$ & $10.46_{-0.84}^{+0.60}$ & $11.00_{-0.51}^{+0.49}$ & $-1.74_{-0.33}^{+0.32}$ & $0.83_{-0.33}^{+0.34}$ & $1.55_{-0.49}^{+0.47}$\\
            \hline
        \end{tabular}       
        \\
        \vspace{1mm}
            \justify{\justify 
            \textsc{Note} --- (1) The stellar bulge mass. (2) The effective radius of the stellar bulge. (3) The \Sersic\ index of the stellar bulge. (4) The mass of the stellar disk. (5) The effective radius of the stellar disk. (6) The stellar-to-DM mass ratio $f_*\equiv M_*/M_h$, where $M_*$ is the total stellar mass and $M_h$ is the DM halo mass. (7) The concentration of the DM halo. (8) The CO-to-H$_2$ conversion factor. The uniform prior limits of parameters are denoted as `(lower, upper)'. The Gaussian priors of parameters are denoted as $\mathrm{\mu\pm\sigma}$. In our MCMC fitting, we set Gaussian priors for $r_\mathrm{e,b}$, $n$ and $r_\mathrm{e,d}$ from \cite{2021ApJ...911...94Z}. We set upper and lower limits for $M_\mathrm{b}$, $M_\mathrm{d}$, $f_*$ and $c$ from literature (\citealt{2021ApJ...911...94Z,2010ApJ...717..379B,2014MNRAS.441.3359D}).\\
            $^a$ \cite{2014MNRAS.441.3359D} suggested a relationship between dark matter halo concentration and dark matter halo mass from numerical simulations, with an uncertainty of $\sim0.1\,$dex. Here we adopt this simulation-driven concentration value $c_\mathrm{fit}$ as the prior knowledge in our dynamical analysis.} 
        \label{table3}
\end{table*}

\begin{figure*}
    \centering
    \includegraphics[width=0.45\linewidth]{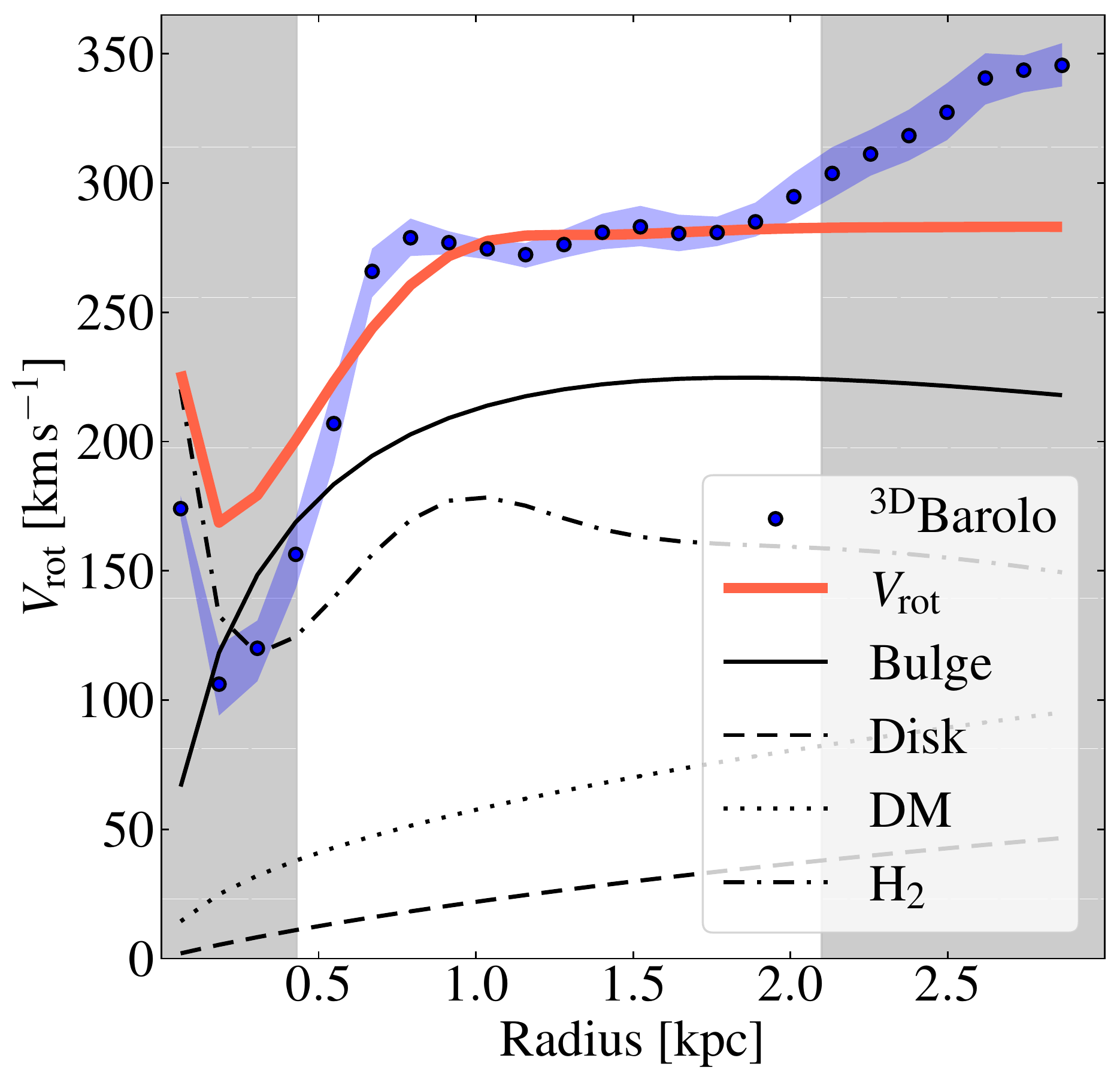}
    \includegraphics[width=0.45\linewidth]{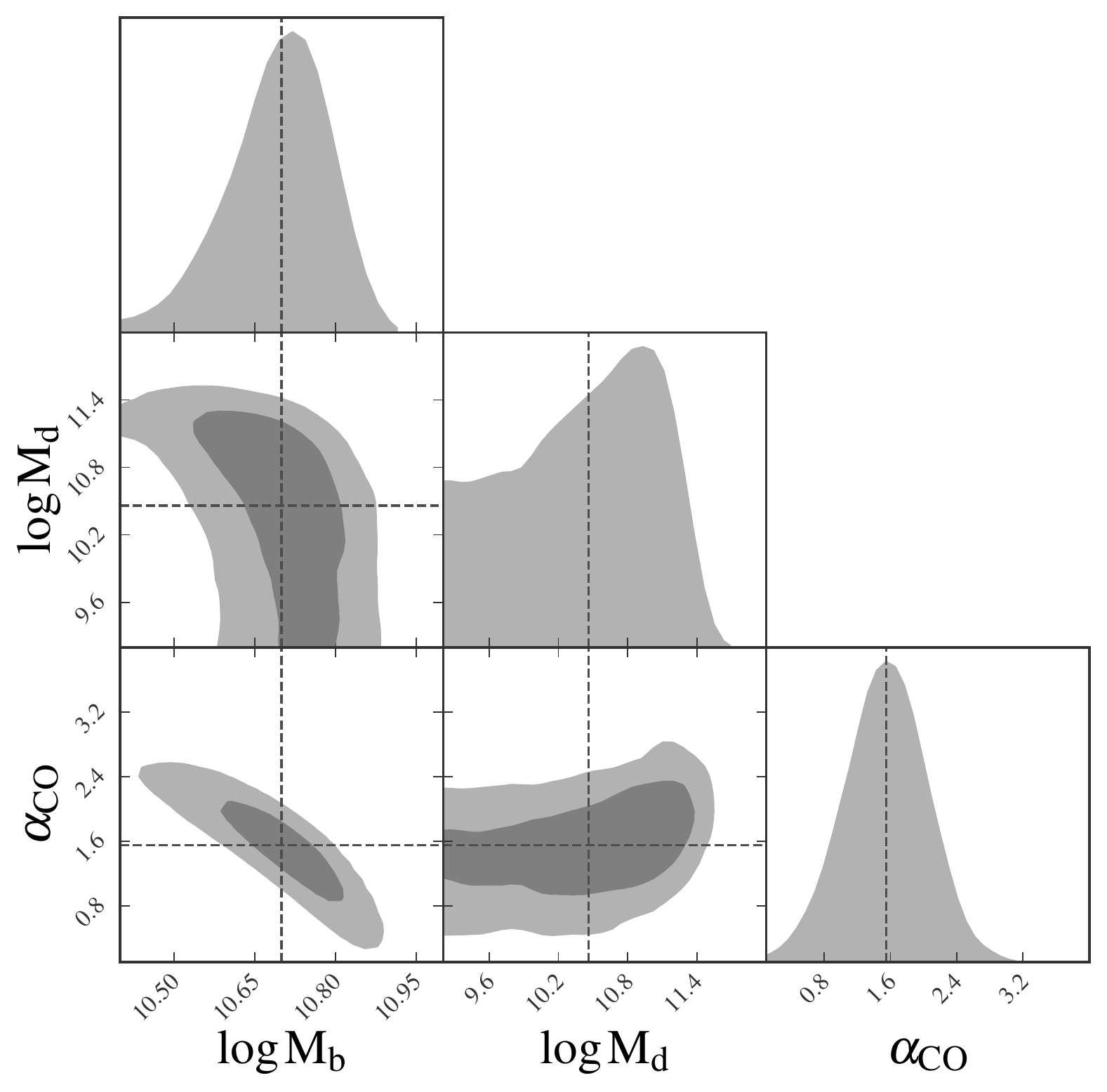}
    \caption{\textbf{Left panel:} Derived rotation curve from \barolo\ and from the best-fitting result. The blue line represents the rotation curve derived from \barolo\ and its surrounding blue shaded region represents the uncertainties. The black solid line shows the rotation curve of the stellar bulge and the dash-dotted line shows the rotation curve of the molecular gas. The dashed line shows the rotation curve of the stellar disk. The dotted line shows the rotation curve of dark matter. The thick red solid line represents the result of the best-fit model rotation velocity. The vertical gray shaded region represents the region within the central synthesized beam area, in which the data points are not used in the fitting. \textbf{Right panel:} The posterior distribution of the stellar bulge mass, stellar disk mass, and the CO-to-H$_2$ conversion factor. The vertical and horizontal dashed lines represent the mean value of each parameter, which are adopted as the best-fitting values and are listed in Table \ref{table3}. The stellar disk mass is poorly constrained, as their contribution is minor in the nuclear region and therefore is heavily degenerate with the stellar bulge component.}
    \label{dynamical_fitting}
\end{figure*}

The rotation curve modeled with \barolo\ provides an independent constraint on the mass distribution within the CO line-emitting region. In this section, we fit the rotation curve with a multi-component dynamical model to investigate the mass budget of the stellar, molecular gas, and dark matter halo. In particular, with knowledge of the dynamical mass measured with the rotation curve and the stellar mass from the HST images \citep{2021ApJ...911...94Z}, we can constrain the mass of molecular gas and estimate the CO-to-\Hmol\ conversion factor, \aco.

Here we model the gas dynamics and fit it to the rotation curve derived from 3D Barolo within the $0.4\sim 2.1$\,kpc radial zone (Figure. \ref{barolo_results}). The inner and outer regions are not considered in the fitting due to the large uncertainties and possible affects from asymmetric structure/spiral arms discussed in Section \ref{3.3}. During the fitting procedure, we assume that the rotation velocity is mainly contributed by four components: stellar bulge, stellar disk, molecular gas disk, and dark matter (DM) halo. We neglect the HI gas component as it is usually much more extended than stars and molecular gas, and thus it only dominates the gas mass on a larger scale (\citealt{2008AJ....136.2563W, 2016MNRAS.460.2143W}). We also do not consider the contribution from the SMBH which has a mass of $9.30_{-1.38}^{+1.26}\times10^{6}\, M_\odot$ \citep{2019ApJ...876..102H} and has neglectable  contribution to the rotation velocity on kpc scale. Thus, the total rotation velocity is calculated as follows:
\begin{align*}
    V_\mathrm{circ,tot}^2=V_\mathrm{bulge}^2+V_\mathrm{disk}^2+V_\mathrm{DM}^2+V_\mathrm{gas}^2,
\end{align*}
where $V_\mathrm{bulge},\,V_\mathrm{disk},\,V_\mathrm{DM}$ and $V_\mathrm{gas}$ are circular velocities contributed by stellar bulge, stellar disk, dark matter halo and molecular gas, respectively. 

\subsection{Circular velocities}
\label{4.1}

For the spherical stellar bulge component, we adopt the deprojected symmetric three-dimensional model from \cite{1997A&A...321..111P}, 
\begin{align}
    \rho(r) = \rho r^{-\alpha}\exp\left(-b_n r^{1/n}\right)
    \label{equationA.B.1}
\end{align}
where $\alpha$ can be estimated as $\alpha=1-1.188/2n + 0.22/4n^2$ (see Equation B7 in \citealt{1997A&A...321..111P}).
A traditional 2D-\Sersic\ profile can be well reproduced by integrating the spatial densities along the line of sight. The circular velocity contributed by the stellar bulge can be written as:
\begin{align*}
    V_\mathrm{bulge}(r)^2 &=\frac{GM(r)}{r},\\
    M(r) &= M_0\frac{\gamma\left[n(3-p), bx^{1/n}\right]}{\Gamma\left[n(3-p)\right]},
\end{align*}
where $r$ is the spatial radius, $M_0$ is the total stellar mass of the bulge. $\Gamma$ and $\gamma$ are gamma and incomplete gamma functions, $x\equiv r/r_e$ is the reduced radius. And when the \Sersic\ index and radius satisfy the relation $0.6<n<10$ and $10^{-2}\leq r/r_e\leq 10^3$, the value of $p$ can be computed as $p=1.0-0.6097/n+0.05563/n^2$ \citep{1997A&A...321..111P}. Three parameters are used to describe the bulge mass distribution, the total mass $M_b$, the effective radius $r_{e,b}$ and the \Sersic\ index $n$.

For the disk component, we use the traditional exponential thin disk model adopted from \cite{2008gady.book.....B}:
\begin{align*}
    V_\mathrm{disk}(r)^2 = 4\pi G \Sigma_0 R_d y^2\left[I_0(y)K_0(y)-I_1(y)K_1(y)\right],
\end{align*}
where $\Sigma_0=M_d/2\pi r_{e,d}^2$ is the surface density, $R_d=r_{e,d}/1.68$ and $y\equiv r/2R_d$. Here, $M_d$ and $r_{e,d}$ are the stellar disk mass and the disk effective radius. $I_i$ and $K_i$ are Bessel functions. Considering the main purpose of our study is to constrain the mass component decomposition, we constrain the $r_{e,b}$ and $r_{e,d}$ from \cite{2021ApJ...911...94Z}. Both parameters have Gaussian priors, with a typical standard deviation of 0.05. The \Sersic\ index for the stellar disk is fixed to 1. 

For the dark matter component, we adopt the simulation-motivated NFW model \citep{1996ApJ...462..563N}, and the circular velocity can be calculated by:
\begin{align*}
    \left[\frac{V_\mathrm{DM}(r)}{V_{\rm vir}}\right]^2=\frac{1}{x}\frac{\ln(1+cx)-(cx)/(1+cx)}{\ln(1+c)-c/(1+c)},
\end{align*}
where $x=r/r_{\rm vir}$ is the radius in units of virial radius, $V_{\rm vir}$ is the virial velocity and $c$ is the halo concentration (see \citealt{1996ApJ...462..563N}, for more details). Considering that our rotation curve only traces the nuclear region, where the contribution from DM is minor and cannot be well constrained, we let the DM parameters satisfy some empirical correlations from numerical simulations. The stellar mass fractions satisfies $-2.3\leq\log (M_*/M_h)\leq-1.3$ for halo mass ranges between $10^{11}M_\odot$ and $10^{13}M_\odot$ \citep{2010ApJ...717..379B}. The concentration follows the function $\log c=a+b\log(M/10^{12}h^{-1}M_\odot)$, and we assume an intrinsic standard deviation of 0.1 dex \citep{2014MNRAS.441.3359D}. Other DM profile (e.g., \citealt{1995ApJ...447L..25B}) is not considered here as the fitting is not sensitive to different assumptions of DM profiles.

We calculate $V_{\rm gas}$ for the molecular gas component following Equation (10) in \cite{2008MNRAS.385.1359N}, which derived rotation curve for an axisymmetric bulge with arbitrary flattening. In this model, the mass density can be written as $\rho=\rho(m)$, with $m=\sqrt{x^2+y^2+(z/q)^2}$ and $q$ is the intrinsic axis ratio of the bulge isodensity surfaces \citep{2008MNRAS.385.1359N}. Models in \cite{2008MNRAS.385.1359N} have four parameters, surface density of gas $\Sigma_\mathrm{g}$, effective radius $r_{e,g}$, \Sersic\ index $n_g$ and the intrinsic axis ratio $q$. Since we know the surface brightness of CO emission from our ALMA observation, we can directly convert the CO line surface brightness to molecular gas surface density with \aco. We adopt an intrinsic axis ratio of $q=H/r_{e}=0.15$ during fitting, by assuming the scale height of molecular gas disk $H\sim150\,\mathrm{pc}$ and $r_{e}\sim1\,\mathrm{kpc}$ from previous studies of gas-rich systems (\citealt{2019ApJ...882....5W,2021ApJ...908..231M}). Based on the line ratio distribution from Figure \ref{line_ratio}, we adopt $R_{21}=0.9$ for the central region, and $R_{21}=0.6$ for the outer region ($R>0.8\,\mathrm{kpc}$) to convert the CO\,(2--1) line intensity to the CO\,(1--0) line.

\subsection{Asymmetric drift correction}
\label{4.2}
The ISM pressure gradients can also provide support to the gas against galaxy self-gravity. This effect needs to be considered and corrected for the rotation curve (asymmetric drift correction; \citealt{2010ApJ...725.2324B,2017ApJ...840...92L}) as:
\begin{align*}
    V_\mathrm{rot}^2 = V_\mathrm{circ,tot}^2+\frac{1}{\rho}\frac{d(\rho\sigma^2)}{d\ln r}
\end{align*}
where $V_\mathrm{circ}$ is the circular velocity derived from the mass model, $V_\mathrm{rot}$ is the observed rotation velocity, and the rightmost term models the effect of asymmetric drift, makeing $V_\mathrm{rot}<V_\mathrm{circ}$. In this term, $\sigma$ is the isotropic velocity dispersion, $\rho$ is the gas density and $r$ is the galactic-to-center radius. Traditionally, $\sigma$ is assumed to be a constant during the application of this asymmetric drift correction (e.g., \citealt{2010ApJ...725.2324B}), however, we find that in this galaxy, $\sigma$ is not constant along the radius (Sec. \ref{3.2}), which means that we cannot directly use this formula.

Considering this, we re-calculate the asymmetric drift correction with our assumption of vertical hydrostatic equilibrium of the molecular gas:
\begin{align*}
    P_\mathrm{ISM} = \mathcal{W},
\end{align*}
where $P_\mathrm{ISM}\propto\rho\sigma^2$ is the ISM turbulent pressure. This assumption is further confirmed in Section \ref{5.3}. We find that the asymmetric drift correction can be written as:
\begin{align}
    V_\mathrm{rot}(r)^2 &= V_\mathrm{circ,tot}(r)^2 + \sigma^2\frac{d \ln \Sigma_\mathrm{g}}{d \ln r} + \sigma^2\frac{d\ln \Sigma_\mathrm{tot}}{d \ln r},
    \label{equation5}
\end{align}
where $\Sigma_\mathrm{g}$ is the gas surface density and $\Sigma_\mathrm{tot}$ is the surface density of the total disk, including gas and stellar component. Since $\sigma$ has been provided by our \barolo\ fitting, and $\Sigma_\mathrm{g}$ and $\Sigma_\mathrm{tot}$ can be derived from $\alpha_{\rm CO}$ and stellar mass during the fitting, we can constrain these parameters with our dynamical model. Although $h_\mathrm{g}$ remains uncertain, we find the variation of this parameter does not affect the asymmetric drift correction significantly, so we assume a constant value of $150\,$pc. We also note that when gravity is dominated by gas or stars, Equation \ref{equation5} reduces to its traditional form in \cite{2010ApJ...725.2324B}.

In general, we have 8 free parameters in our fitting: the mass of stellar bulge and stellar disk, the \Sersic\ index of the stellar bulge, the effective radius of the stellar bulge and the stellar disk, the dark matter halo mass and concentration, and the CO-to-H$_2$ conversion factor. 

We use the \texttt{emcee} to determine the best-fit values. In order to minimize the free parameters and avoid degeneracy, we set prior constraints following the available stellar galaxy morphology models for the host galaxy, as well as the relation between the DM halo and stellar mass content. We set Gaussian priors for the effective radius of the stellar bulge and stellar disk, and the \Sersic\ index of bulge following the fitting results and uncertainties from the $B$ and $I$ band HST image modeling \citep{2021ApJ...911...94Z}. The prior constraints and posterior derived values are listed in Table \ref{table3}.

\subsection{Best-fit dynamical modeling and results}
\label{4.3}

We start the MCMC sampling by using 400 walkers, with 1000 steps after a burn-in of 400 steps. We then adopt the 50th percentile of samples as the best-fit values, and estimate uncertainties using 16th and 84th percentiles of the samples. The best-fit results and the posterior distribution of bulge mass, disk mass and \aco\ are shown as Figure \ref{dynamical_fitting}.

We find that the geometry parameters (effective radius and \Sersic\ index) are the same as their prior constraints, which suggests that the rotation curve cannot provide enough information to determine all of the parameters. We still use these prior constraints rather than fix them in order to take into account the uncertainties of these parameters from \cite{2021ApJ...911...94Z}. We find that the masses of the stellar bulge and stellar disk are consistent with that in \cite{2021ApJ...911...94Z}, while the stellar disk mass has quite large uncertainty. Such large uncertainties are also found in the posterior distribution of the stellar fraction and concentration parameters of dark matter halo. The large uncertainties imply that they are hardly constrained in our dynamical model; the gravitational potential of the region traced by CO\,(2--1) is dominated by stellar bulge and molecular gas. Other components only have minor contributions, and the fitting results are heavily affected by the uncertainties of the stellar bulge and molecular gas masses. We also test the fitting results with different initial setups, and in most cases, the posterior probability distributions are consistent with each other (see more details in Appendix \ref{C}). 

We derive \aco\ $=1.55_{-0.49}^{+0.47}$\,\acounit from our dynamical method. This value is between the MW-like [$\alpha_{\rm CO} \approx 4.3\,$\acounit] and ULIRG-like value [$\alpha_{\rm CO} \approx 0.8\,$\acounit; \citealt{1998ApJ...507..615D,2013ARA&A..51..207B}]. This value is $\sim 2$ times smaller than that in nearby star-forming galaxies \citep{2013ApJ...777....5S}. The \aco\ value we derived here is only valid within the $\sim2\,$kpc region of the quasar host galaxy, where the high gas surface density and star formation rate surface density suggest a nuclear starburst (see Section \ref{5.2}). The current CO\,(2--1) data cannot trace the molecular gas in the extended galactic disk and spiral arm region where a higher \aco\ value may remain more appropriate to estimate the molecular gas mass. We also check whether this \aco\ value is reasonable given some theoretical prescriptions (e.g., \citealt{2013ARA&A..51..207B}). \cite{2013ARA&A..51..207B} indicated that \aco\ could have large variations depending on metallicity and gas surface density. However, we note that I\,Zw\,1 presents a metallicity of $\log \left(\mathrm{O/H}\right)=8.77$ by adopting the $M_* - Z$ relationship obtained for SDSS galaxies with the \cite{2004MNRAS.348L..59P} calibration (\citealt{2008ApJ...681.1183K}), which is close to solar metallicity. Therefore, we do not expect any significant \aco\ variation due to high metallicity. We estimate an \aco\ value of $\sim 1.9\,$\acounit\ by solving the \aco--$\Sigma_\mathrm{mol}$ relation of \cite{2013ARA&A..51..207B}. This value is consistent with our dynamical \aco\ value considering the uncertainties.

By adopting this new \aco, we estimate a total cold molecular gas mass at a value of $\log M_\mathrm{H_2}/M_\odot=9.94_{-0.31}^{+0.18}$, and the gas fraction is $f_\mathrm{gas}= 0.10_{-0.08}^{+0.12}$. The value of the gas fraction is similar to that in inactive star-forming galaxies and hard X-ray selected AGN host galaxies (\citealt{2020ApJ...899..112S,2021ApJS..252...29K}), and is smaller than that in local LIRGs by a factor of $\sim$2 \citep{2016ApJ...825..128L}.

Using our dynamical method, we firstly investigate the \aco\ value in this quasar host galaxy, and find that the value of \aco\ is between that in ULIRGs and in the MW \citep{2013ARA&A..51..207B,2020A&A...643A..78M}. In the rest part of this work, we estimate the molecular gas mass by adopting the median CO-to-\Hmol\ conversion factor value derived from our best dynamical model. 

\section{Discussion}
\label{sec5}
\subsection{Distribution of the CO\,(2--1)-to-CO\,(1--0) line ratio}
\label{5.1}

\begin{figure}
    \centering
    \includegraphics[width=\linewidth]{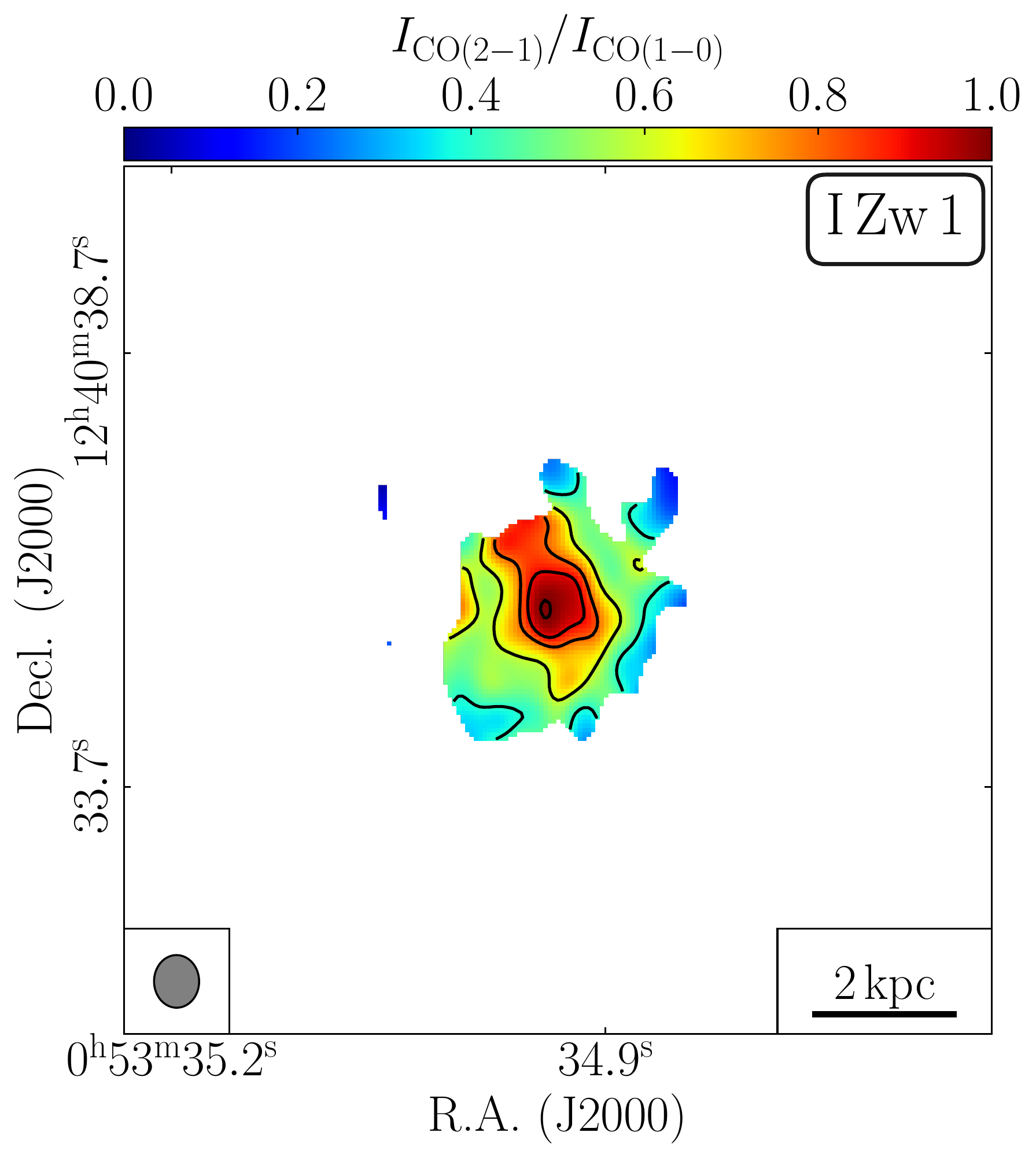}
    \caption{Surface brightness ratio between CO\,(2--1) and CO\,(1--0) in I\,Zw\,1. The contour level corresponds to $[0.4, 0.6, 0.8, 0.9, 1.0]$. The ellipse in bottom left represents beam size of $0.61''\times 0.52''$ for the CO\,(1--0) observation.}
    \label{line_ratio}
\end{figure}

A CO\,(2--1) to CO\,(1--0) line luminosity ratio of $R_{21}=0.63\pm0.02$ was reported in \cite{2020ApJS..247...15S} based on previous ACA measurements of the total gas content, which is within the typical range for subthermal CO-excited molecular gas in galactic disks ($R_{21}< 0.8$; \citealt{2013AJ....146...19L, 2015AAS...22514125R, 2017ApJS..233...22S}). Here we report the surface brightness ratio distribution estimated from resolved ALMA images of CO\,(2--1) and CO\,(1--0) that are shown in Section \ref{sec2}. We smooth the CO\,(2--1) line image with CASA task \texttt{imsmooth} to match the angular resolution of the CO\,(1--0) data. We estimate the surface brightness ratio with \texttt{immath} within the region where both signal-to-noise ratios are larger than 5. The emission line ratio map is shown in Figure \ref{line_ratio}.

The line intensity ratio is close to 1 within the radius of $\sim$1\,kpc in the quasar host galaxy, suggesting that the molecular gas in the central region is optically thick and thermalized. The high $R_{21}$ value in the nuclear region is consistent with the previous result reported by \cite{2004ApJ...609...85S} based on Berkeley-Illinois Maryland Association (BIMA) and Plateau de Bu Interferometer (PdBI) observations at a lower angular resolution of $\sim$0.7$''$ for CO(1--0) and of $\sim 0.9''$ for CO(2--1) data. A similar emission line ratio distribution with a higher value toward the center is commonly found in nearby spiral galaxies (\citealt{1992A&A...264..433B, 2021MNRAS.504.3221D, 2021PASJ...73..257Y}), local IR luminous galaxies \citep{2012MNRAS.426.2601P} and high redshift galaxies (\citealt{2013ARA&A..51..105C, 2015A&A...577A..46D}). 

In the outer disk region, the emission line ratio is relatively low ($\lesssim 0.6$). It is likely that at larger radii, the molecular gas becomes subthermally excited \citep{2017MNRAS.470.1570H} or has a lower temperature \citep{1992A&A...264..433B}.

\subsection{Star formation law of the nuclear region}
\label{5.2}

\begin{figure}
    \centering
    \includegraphics[width=\linewidth]{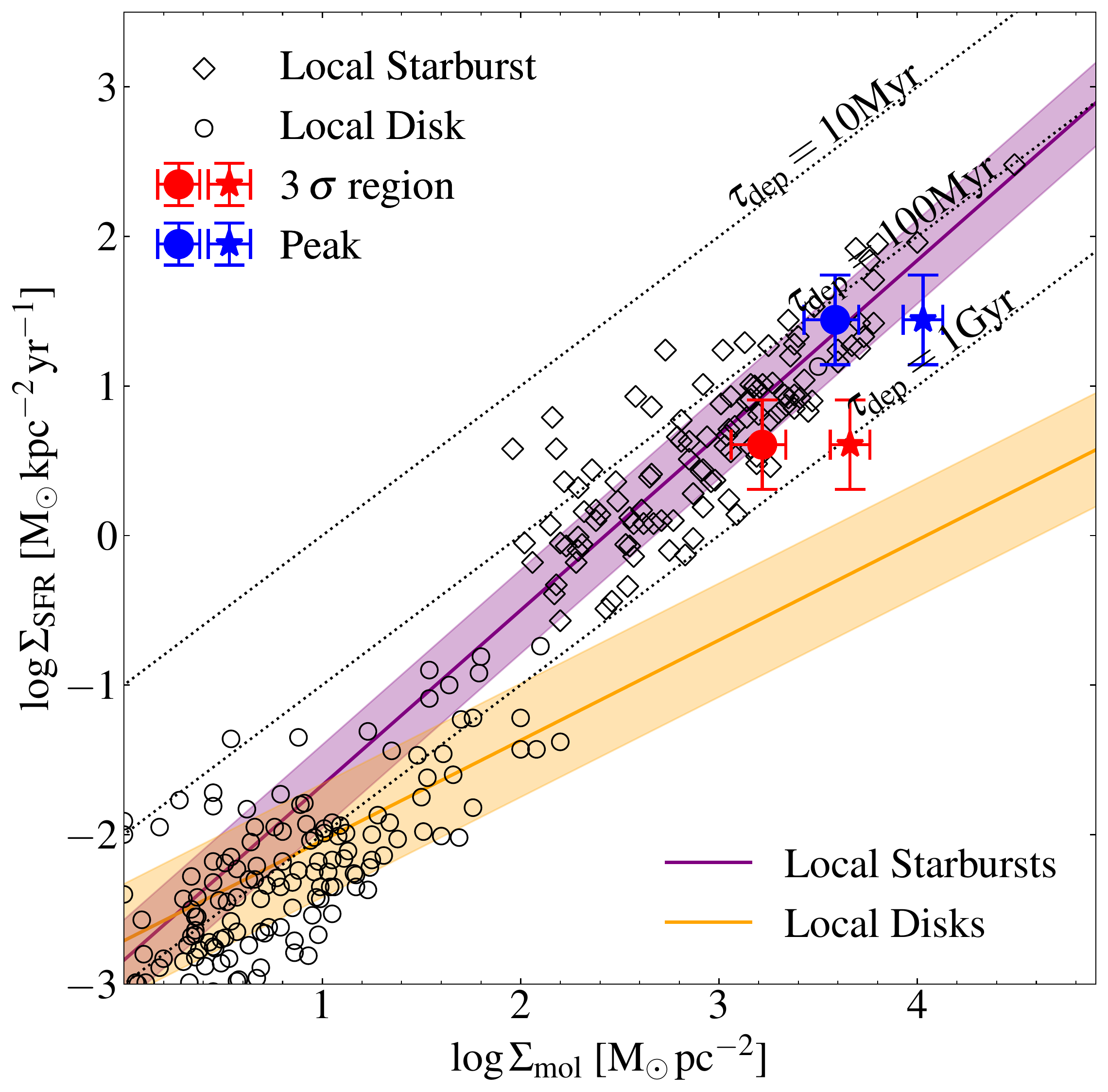}
    \caption{The surface density of molecular gas vs. surface density of SFR in the nuclear of I\,Zw\,1. The open markers represent star-forming (circles; \citealt{2019ApJ...872...16D}) and starburst galaxies (diamonds; \citealt{2021ApJ...908...61K}) in the local universe. The orange and purple solid lines represent the KS-law for local star-forming and starburst galaxies \citep{2019ApJ...872...16D,2021ApJ...908...61K}, whose surrounding shaded region represents the scatter on a order of $\sim 0.3$\,dex. The filled red point shows the mean surface density measured within the 3 $\sigma$ contour region of the continuum map (Figure \ref{continuum}) while the filled blue point gives the peak values of the molecular gas and SFR surface densities measured within the central beam. The filled stars represent the gas surface density if a MW-like \aco\ is adopted. The dotted lines show the trends with gas depetion timescales $\tau_{\rm dep}=10$\,Myr, 100\,Myr, and 1\,Gyr.}
    \label{KS_law}
\end{figure}

We check the Kennicutt-Schmidt relation \citep{1998ApJ...498..541K} in the nuclear region of I\,Zw\,1 based on the CO\,(2--1) line and continuum maps and the new \aco\ value of $\sim 1.5\,$\acounit\ derived from our dynamical modeling fitting. 

The ALMA continuum image reveals an 1.3\,mm continuum flux density of $\sim1.1$\,mJy from the central 3 $\sigma$ contour region. \cite{2022arXiv221205295M} decomposed the SED of I\,ZW\,1. They estimated that the AGN contribution, including the non-thermal synchrotron emission extrapolated from the radio bands and the thermal free-free emission, contribute about 28\% ($\sim$0.30\,mJy) of the ALMA continuum, and the remaining 72\% is likely to be from the thermal dust heated by nuclear star formation. Based on this decomposition of the ALMA continuum, they calculated a nuclear star formation rate of 5.43\,$M_\odot\,\rm yr^{-1}$.

The AGN contribution to the millimeter dust continuum emission could also be estimated and removed based on the empirical luminosity relations. \cite{2022ApJ...938...87K} presented a relationship between the rest-frame 1.3\,mm-wave ($\nu L_\mathrm{\nu,\,mm}$) and 2-10 keV X-ray luminosities ($L_{\rm 2-10\,keV}$) for AGNs (Table 1 in \citealt{2022ApJ...938...87K}). Based on this relation and adopting the 2-10\,keV luminosity of I\,ZW\,1 from \citep{2005A&A...432...15P}, we estimate an AGN contribution to the 1.3\,mm continuum flux density of $0.35_{-0.23}^{+0.65}$\,mJy. This flux density has been corrected to the observing frame assuming a mm-wave spectral index of 0.5 ($S_{\rm 1.3\,mm} \propto \nu^{-0.5}$). This value is consistent with that derived from the synchrotron and free-free components in the  SED decomposition, considering the large uncertainty of 0.45 dex of the $\nu L_{\rm \nu,\,mm}$ --- $L_{\rm 2-10\, keV}$ relation. Therefore, we adopt the nuclear star formation rate of 5.43\,$M_\odot\,\rm yr^{-1}$ from \cite{2022arXiv221205295M} in the analysis here.

The face-on size of the star-forming region is estimated by $A=S/\cos i$, where $S$ is the area within the 3$\sigma$ contour region of continuum map, and $i$ is the inclination angle. Thus the mean surface density of star formation rate in the nuclear region can be estimated by $\Sigma_\mathrm{SFR}=\mathrm{SFR}/A$.

We then estimate the mean molecular gas surface density using the CO\,(2--1) flux in the aformentioned region. By assuming $R_{21}$ following the CO emission line ratio map presented in Figure \ref{line_ratio}, we estimate the CO\,(1--0) emission line flux in this region. Finally we estimate the gas surface density with $\Sigma_\mathrm{mol}=\alpha_\mathrm{CO}\times L'_\mathrm{CO\,(1-0)}/A$, where \aco$=1.5\,$\acounit is the CO-to-\Hmol\ conversion factor and $L'_\mathrm{CO\,(1-0)}$ is the CO\,(1--0) luminosity in the nuclear region. We also present the estimation of molecular gas surface density by adopting an \aco\ value of $4.3\,M_\odot\mathrm{\left(K\,km\,s^{-1}\,pc^2\right)^{-1}}$, which is the typical value of the MW-like galaxy, for comparison \citep{2013ARA&A..51..207B}.

We compare the derived surface densities of SFR and of molecular gas mass in the plot of the KS-relation in Figure \ref{KS_law}. The mean SFR and molecular gas surface densities in the nuclear region (filled blue circle) of I\,Zw\,1 is comparable to the typical values of starburst galaxies (open diamonds in Figure \ref{KS_law}; \citealt{2012A&A...539A...8G, 2021ApJ...908...61K}), and larger than that of the normal star-forming galaxies (open circles in Figure \ref{KS_law}; \citealt{2013AJ....146...19L, 2019ApJ...872...16D}). We also investigate the surface densities of SFR and molecular gas for the central peak (filled red circle), and find that this data point locates well within the starburst source region. The gas depletion timescale is derived using $\tau_{\rm dep}=\Sigma_{\rm mol}/\Sigma_{\rm SFR}$. We found $\tau_{\rm dep} \sim 300\,$Myr, which is close to that of local starburst systems. We also present the gas surface densities adopting $\alpha_{\rm CO} = 4.3\,$\acounit\ from \cite{2013ARA&A..51..207B}, which are shown as filled red and blue stars in Figure \ref{KS_law} and locate below the KS-relation for starburst systems.

We find that this quasar host galaxy has enhanced star-forming activity in its central $\lesssim 500\,$pc region. The starburst activity suggests that AGN feedback plays a minor role in stopping ongoing star formation, and a positive influence can also be plausible. Assuming a MW-like \aco\ increases the molecular gas surface density significantly, while the data points are still close to ULIRG-like KS relation, and are still well above the KS relation for local star-forming galaxies (Figure \ref{KS_law}). Nuclear starbursts were also found in other low-$z$ quasars \citep{2004A&A...423L..13C,2006ApJ...649...79S,2022arXiv221205295M}.

\subsection{Does AGN perturb the cold molecular gas?}
\label{5.3}

\begin{figure}
    \centering
    \includegraphics[width=\linewidth]{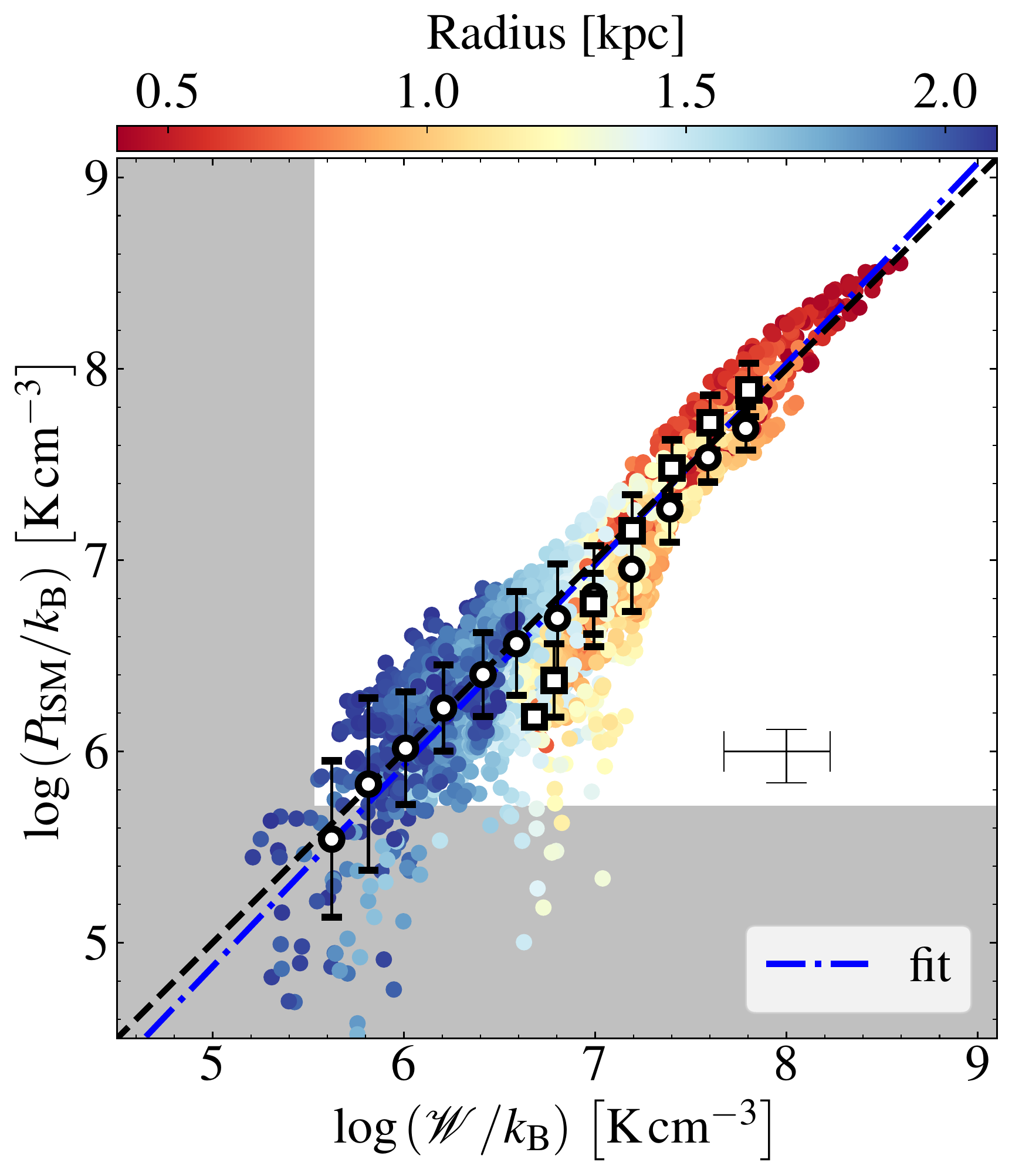}
    \caption{The ISM turbulent pressure ($P_\mathrm{ISM}$) as a function of ISM weight ($\mathcal{W}$). Data points represent pixels in CO\,(2--1) map and are color-coded by the distance between pixels and the center of this galaxy, respectively. The open squares and circles represent the mean ISM turbulent pressure of each ISM weight bin, and denote data points in group (2) and group (3) as described in Section \ref{5.3}. The gray shaded region represents the lower limits of both pressure by considering the CO\,(2--1) detection limits. The dashed line denotes equality. The blue dash-dotted lines represent the best-fitting power-law results. The typical uncertainty is plotted in the lower right corner. This uncertainty is dominated by \aco, which is $\sim0.2$\,dex.}
    \label{equilibrium}
\end{figure}

We measure an intrinsically large velocity-dispersion in the galactic nucleus from our kinematic analysis in Section \ref{3.2}, which is $3 \sim 4$ times higher than the values measured in the outer region with radii $>0.8\,$kpc. Such a large velocity-dispersion indicates that the molecular gas in this central 1\,kpc region has a large turbulent energy. There are several possible mechanisms that may contribute to the turbulent ISM. This large velocity-dispersion might be related to the high gas surface density, since the ISM turbulent pressure ($P_\mathrm{ISM}$) should be in equilibrium with the weight ($\mathcal{W}$) of the ISM (\citealt{2020ApJ...901L...8S, 2022ApJ...936..137O}). The enormous energy released by central AGN could also perturb the ISM. The central starburst may also enhance the gas velocity dispersion by means of stellar feedback. To check whether such large velocity-dispersion originates from the AGN and stellar feedback, we try to identify whether there is an excess in the ISM turbulent pressure ($P_\mathrm{ISM}$), comparing to the weight ($\mathcal{W}$) of the ISM. Any excess ISM turbulent pressure should represent the energy released by central AGN or starburst.

The weight of the ISM can be expressed as follows \citep{2022ApJ...936..137O}:
\begin{align}
    \mathcal{W}=\pi G \Sigma_\mathrm{g}^2/2 + 4\pi \zeta_\mathrm{d} G \Sigma_\mathrm{g} \rho_\mathrm{sd} h_\mathrm{g}.
    \label{weight_of_ISM}
\end{align}

The first term is the weight due to the self-gravity of the ISM disk (\citealt{1942ApJ....95..329S,1989ApJ...338..178E}). Here the surface density of ISM should have at least two components in principle, i.e., $\Sigma_\mathrm{g}=\Sigma_\mathrm{H_2}+\Sigma_\mathrm{HI}$. However, we note that the atomic gas is negligible in the galaxy center. Therefore we replace $\Sigma_\mathrm{g}$ with $\Sigma_\mathrm{H_2}$ in the remainder context. The second term is the weight of the ISM due to external gravity including the stellar component and dark matter halo. The numerical value of $\zeta$ depends on, but not sensitively to, the geometric distribution of gas disk, thus can be assumed as a constant of $\sim 1/3$(see Equation 6 of \citealt{2010ApJ...721..975O}), and $\rho_\mathrm{sd}$ is the external density. In the galaxy center where the external gravitational potential is dominated by the stellar bulge, $\rho_\mathrm{sd}$ could be estimated using the bulge mass density of $\rho_{\rm b}$. This term also accounts for the half-thickness of the gas disk, $h_\mathrm{g}$, whose typical value is about $100 - 200$\,pc \citep{2019ApJ...882....5W}.

The ISM turbulent pressure at the midplane is defined by the difference in the total vertical momentum flux across the gas layer and thus can be expressed as:
\begin{align}
    P_\mathrm{ISM} = \rho_\mathrm{mid}\sigma_\mathrm{g}^2(1+\alpha+\beta)=\frac{\Sigma_\mathrm{g}}{2h_\mathrm{g}}\sigma_\mathrm{g}^2(1+\alpha+\beta),
    \label{total_pressure}
\end{align}
where $\sigma_\mathrm{g}$ is the velocity dispersion of molecular gas and can be measured from the CO emission line. The parameters $\alpha\approx0.3$ and $\beta\approx0.0$ are the factors accounting for vertical-magnetic and cosmic-ray pressures contribution (\citealt{2015ApJ...815...67K, 2019ApJ...882....5W}).

As was pointed out in \cite{2022ApJ...936..137O}, the vertical hydrostatic equilibrium requests that the ISM weight $\mathcal{W}$ must be equal to the ISM turbulent pressure $P_\mathrm{ISM}$,
\begin{align}
    \mathcal{W}=P_\mathrm{ISM}.
    \label{equation4}
\end{align}

We examine the relationship between the ISM turbulent pressure and the weight of ISM in this quasar host galaxy, pixel-by-pixel. We estimate the pixel-wise velocity dispersion using the following methods. In the first step, we generate the moment 2 map by applying a blanking mask using python package \texttt{maskmoment}.\footnote{\url{https://github.com/tonywong94/maskmoment}} The mask was created by starting at 5$\sigma$ peaks in the cube, and expanding down to the surrounding 2$\sigma$ contour. The final moment maps were refered as `dilated-mask' moment maps in \cite{2017ApJ...846..159B}. This dilated moment 2 map can simultaneously capture low-level signal and avoid noise, and can well reproduce the observed velocity dispersion. In the second step, we build a rotating disk model using \barolo\, to simulate the velocity dispersion that is solely caused by beam-smearing effect. The model has the same CO\,(2--1) line intensity and rotation velocity with that from the I\,Zw\,1 CO\,(2--1) data (Section \ref{3.2}), but the velocity dispersion of the model is set to be zero. Then the model is convolved with the synthesized beam of I\,Zw\,1 CO\,(2--1) data. We generate the simulated velocity dispersion from this model with a 2$\sigma$ cutoff threshold. This simulated velocity dispersion is then removed from the dilated moment 2 map in quadrature to generate the intrinsic $\sigma_{\rm g}$ map \citep{2018ApJ...860...92L}.

Based on the line ratio map discussed in Section \ref{5.1}, we estimate the CO\,(1--0) line surface brightness assuming $R_{21}=0.9$ at $R<0.8\,$kpc, and $R_{21}=0.62$ at $R>0.8\,$kpc \citep{2020ApJS..247...15S}. We then estimate the molecular gas mass by adopting $\alpha_\mathrm{CO}=1.55\,$\acounit\ from the dynamical modeling as shown in Section \ref{4.3}. The $\rho_\mathrm{sd}$ is estimated from dynamical modeling and follow Equation (\ref{equationA.B.1}). We estimate these two pressures by assuming a constant gas disk scale height of $h_\mathrm{g}=150\,$pc, which is a typical value of nearby ULIRG and starburst systems (\citealt{2019ApJ...882....5W,2020A&A...643A..78M}). The relation between $P_\mathrm{ISM}$ and $\mathcal{W}$ are plotted in Figure \ref{equilibrium}. We manually separate pixels into four groups depending on their radii: (1) $R<0.4\,$kpc, (2) $0.4\,\mathrm{kpc}< R< 0.8$\,kpc, (3) $0.8\,\mathrm{kpc}< R< 2.1$\,kpc, and (4) $R>2.1\,$kpc. We do not include data from group (4) as the S/N of the CO\,(2--1) line in this region is too low. Also, we avoid presenting the data points from group (1) due to severe beam-smearing effect.

Figure \ref{equilibrium} shows the relationship between $P_\mathrm{ISM}$ and $\mathcal{W}$, spanning three orders of magnitude. The color of each data point represents the distance between each pixel and the galaxy center. In order to highlight the difference in group (2) and (3), we calculate the mean and scatter trends for each group in 0.2\,dex wide bins of fixed ISM weight.

We fit the relation between $P_\mathrm{ISM}$ and $\mathcal{W}$ in logarithmic space using the python package \texttt{linmix} \citep{2007ApJ...665.1489K}. This yields the best fitting power-law relations (blue dash-dotted line in Figure \ref{equilibrium}):
\begin{align*}
    \log \left(\frac{P_\mathrm{ISM}}{k_\mathrm{B}\,\mathrm{K\,cm^{-3}}}\right) &= \left(-0.38_{-0.56}^{+0.58}\right) \\
    &+ \left(1.05_{-0.07}^{+0.08}\right) \log \left(\frac{\mathcal{W}}{k_\mathrm{B}\,\mathrm{K\,cm^{-3}}}\right).
\end{align*}
The best fitting result is consistent with the equality relation (black dashed line in Figure \ref{equilibrium}) considering the uncertainty. This result suggests that the origin of the high turbulent energy of the cold molecular gas (with $\sigma\sim100\,$\kms) can be explained by the self-gravity of the galaxy alone.

\subsection{The lack of negative AGN feedback}
\label{5.4}

The ALMA CO\,(2--1) image reveals that the molecular gas in the host galaxy of I\,Zw\,1 is centrally concentrated with a high surface density in the central kpc region where intense star formation is taking place. This result is contradictory to the scenario in which AGN feedback can efficiently blow out the star-forming gas from the nuclear region, and results in the depletion of cold gas in galaxy center (\citealt{2011ApJ...729L..27R, 2021MNRAS.505L..46E}). In addition, there is no evidence of AGN-driven outflow in the nuclear region like other AGN host galaxies (e.g., \citealt{2020ApJ...890...29F}). As an alternative, we find an enhancement of gas velocity dispersion in the nuclear region, which indicates that the nuclear gas is dynamically hot compared with gas in the circumnuclear disk. However, we find that ISM turbulent pressure is in equilibrium with the weight of ISM, suggesting that the kinematics of molecular gas could be regulated by the host galaxy's self-gravity. The large velocity dispersion is naturally required to satisfy the hydrostatic equilibrium. There is no external energy budgets/pressure, e.g., from AGN feedback, that is expelling the cold gas from the galaxy center

So far, \cite{2020ApJS..247...15S} reported that I\,Zw\,1 is a CO luminous system and there is no evidence of galactic scale molecular gas outflows. \citep{2022arXiv220903380L} also report non-detection of molecular gas outflow from this object based on the ALMA high resolution CO (2-1) data. \cite{2021ApJ...908..231M} found that the molecular gas in this galaxy is centrally concentrated, and rotating in a disk with negligible non-circular motions. Moreover, the continuum map reveals a centrally enhanced star formation (see also \citealt{2022arXiv221205295M}) which also argue against the suppression of star formation from AGN feedback. From our kinematic and dynamical analysis, we find no evidence of AGN driven outflow or external gas energy budget. In addition, ionized gas components with  high-velocity dispersions were detected in some nearby quasar host galaxies from recent optical IFU data (\citealt{2019A&A...627A..53H,2022A&A...659A.123S}). However, \cite{2022ApJ...935...72M} reported that the kinetic energy of these gas components with high-velocity dispersions is only $\lesssim 0.1$\% of the AGN bolometric luminosities. This suggests that only a negligible percentage of the AGN power is coupled to the ISM. All these results suggest a lack of negative AGN o the cold molecular gas and star formation in the quasar host galaxy.

\section{Conclusions}
\label{sec6}
We present a study of CO\,(2--1) line emission in the nuclear region of I\,Zw\,1 based on ALMA observations. A combination of all available data from the ALMA archive resolves the CO source on 0.36$''$ scale with a spectral sensitivity of 0.28\,\mjybeam per channel. In the central 1\,kpc region, molecular gas forms a high-density bar-like structure, which has a different position angle compared to that of the main disk.
\begin{enumerate}
    \item[$\bullet$] With \barolo\ fitting, we obtain the intrinsic rotation velocity and velocity dispersion as a function of radius. This galaxy is a rotation-dominated system, similar to other star-forming galaxies in the local universe. The mean rotation velocity to dispersion ratio is about nine, which suggests that the molecular gas forms a cold disk. Meanwhile, the fitting results from the \barolo\ model suggests an enhancement of velocity dispersion in the central sub-kpc scale region. We check the velocity field carefully and find that the pure beam-smearing effect cannot lead to such a large velocity dispersion. The velocity dispersion of the molecular gas in the central region of nuclear disk is intrinsically $\sim 3$ times higher compared to that in the disk region. 
    \item[$\bullet$] The map of the emission line ratio between two CO emission lines represents a clear gradient of $R_{21}$ along the radius. The central value is close to the theoretical prediction under the assumption of thermalized, optically thick ISM conditions. In contrast, the main circumnuclear disk has relatively lower values.
    \item[$\bullet$] We fit the rotation curve of the molecular gas disk and constrain the mass budget of the quasar host galaxy using a dynamical model. We take into account the constraints on gas distribution from the ALMA CO data and stellar morphology/mass from the HST image, and fit the CO-to-\Hmol\ conversion factor. We find a best-fit $\alpha_{\rm CO} = 1.55_{-0.49}^{+0.47} \, M_\odot \,($K\,km\,s$^{-1}$\,pc$^2)^{-1}$, which is between the ULIRG-like and MW-like \aco\ value [$\alpha_{\rm CO,ULIRG}\approx 0.8\,$\acounit, $\alpha_{\rm CO,MW}\approx 4.3\,$\acounit]. 
    \item[$\bullet$] We check the star formation rate and molecular gas surface densities in the central region, finding that the star formation activity follows the Kennicutt-Schmidt relation of local starburst galaxies, which suggests a nuclear starburst activity. 
    \item[$\bullet$] By comparing the ISM turbulent pressure ($P_\mathrm{ISM}$) and the weight of the ISM ($\mathcal{W}$), we find these two parameters are almost equal to each other. The ISM turbulent pressure is in equilibrium with galaxy gravity, which suggests molecular gas in this galaxy is regulated by its self-gravity, and there is no external energy budgets that are exploring the cold gas. This result indicates that the central AGN, though luminous in the optical, is unlikely to introduce extra pressure to the molecular gas in the nuclear region.
\end{enumerate}

\begin{acknowledgments}
We acknowledge supported by the National Science Foundation of China (11991052, 11721303, 12173002, 12011540375) and the China Manned Space Project (CMS-CSST-2021-A04, CMS-CSST-2021-A06); ANID grants PIA ACT172033 (E.T.), Basal-CATA PFB-06\/2007 and AFB170002 grants (E.T., F.E.B.), FONDECYT Regular 1160999, 1190818 (E.T., F.E.B.), and 1200495 (E.T., F.E.B.), and Millennium Science Initiative ICN12\_009 (F.E.B.). This paper makes use of the following ALMA data: ADS/JAO.ALMA\#2015.1.01147.S, \#2017.1.00297.S, \#2018.1.00006.S, \#2018.1.00699.S. ALMA is a partnership of ESO (representing its member states), NSF (USA) and NINS (Japan), together with NRC (Canada), MOST and ASIAA (Taiwan), and KASI (Republic of Korea), in cooperation with the Republic of Chile. The Joint ALMA Observatory is operated by ESO, AUI/NRAO and NAOJ.
\end{acknowledgments}
\facilities{ALMA}
\software{astropy \citep{2013A&A...558A..33A}; CASA \citep{2007ASPC..376..127M}; emcee \citep{2013PASP..125..306F}; numpy \citep{2011CSE....13b..22V}; scipy \citep{2020NatMe..17..261V}}

\appendix
\section{Testing the gas velocity dispersion with simulated data}
\label{A}
\begin{figure*}
    \centering
    \includegraphics[width=\linewidth]{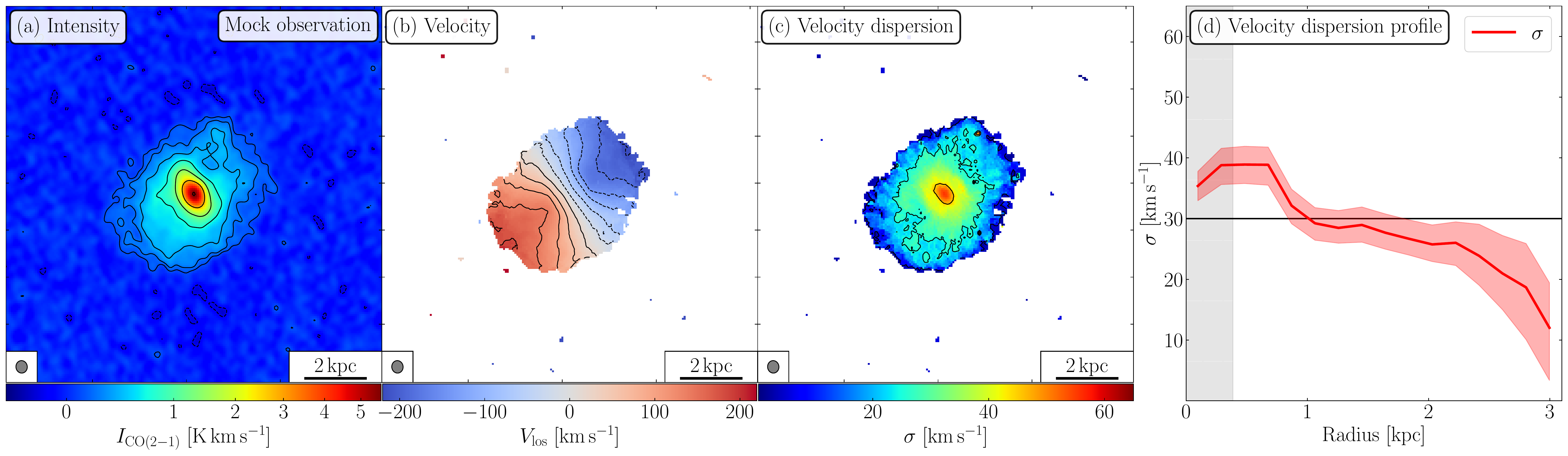}
    \caption{Panel (a), (b), and (c) show the velocity-integrated intensity map, the flux-weighted line-of-sight velocity map, and the velocity dispersion map of simulated data. The synthesized beam is shown as a gray ellipse in the bottom left corner of each panel. The scale bar is shown in the bottom right corner of each panel. Contours are as same as those in Figure \ref{intensity_fitting}. Panel (d) represents the velocity dispersion extracted from the mock observation through \barolo. The red curve represents the best-fitting velocity dispersion from \barolo. And the red shaded region represents the uncertainties. The horizontal line represents the input velocity dispersion of simulated data.  The synthesized beam ($0.33''\times0.30''$) is plotted at the lower left corner of firet 3 panels.}
    \label{intensity_fitting0}
\end{figure*}

We build a mock observational data cube to test whether \barolo\ is able to reproduce the intrinsic gas velocity dispersion with the reduction of beam smearing effect. We build the rotating disk model with \barolo\ \texttt{galmod} task. The model has the same CO\,(2--1) line intensity and rotation velocity with that from the I\,Zw\,1 CO\,(2--1) data. But the model has a constant velocity dispersion ($\sigma=30\,\mathrm{km\,s^{-1}}$) along all radii. We then simulate the visibility data with CASA task \texttt{simobserve}, and and image and clean this simulated visibility using the same procedure mentioned in Section \ref{sec2}. We adjust the total integration time and ALMA configuration to obtain a similar signal-to-noise ratio and angular resolution (the angular resolution of mock observation is $0.33''\times0.30''$). We then use \barolo\ to fit the simulated data cube and the result is shown in Figure \ref{intensity_fitting0}. We can see that, if a gaseous rotating disk has a constant velocity dispersion with a value of 30\,\kms\ along all radii, the beam smearing effect can boost the velocity dispersion up to $\sim60\,$\kms\ in its center. With \barolo\ analysis, although we cannot completely reduce the beam smearing effect, the velocity dispersion has an error of $\lesssim30\%$ in the central region ($R\lesssim1\,\mathrm{kpc}$). This result indicates that the centrally enhanced gas velocity dispersion ($\sigma\gtrsim100\,\mathrm{km\,s^{-1}}$) that is found in the host galaxy of I\,Zw\,1 may not be solely produced by beam smearing effect. Molecular gas in the center of this galaxy should have a very large velocity dispersion ($\sigma\gtrsim100\,\mathrm{km\,s^{-1}}$) intrinsically. We also find that the velocity dispersion decreases at a large radius. This result is caused by the decreasing of S/N of the simulated CO emission at the disk edge.

\section{Dynamical models with different prior constraints}
\label{C}

\begin{table*}
    \centering
    \caption{Constraints and results of dynamical parameters}
    \begin{tabular}{cccccccc}
        \hline
        \hline
        Cases & \multicolumn{3}{c}{\textbf{Prior}} & & \multicolumn{3}{c}{\textbf{Posterior}}\\
         \cline{2-4} 
         \cline{6-8}
         & $\log M_\mathrm{b}$ & $\log M_\mathrm{d}$ & $\alpha_\mathrm{CO}$ & & $\log M_\mathrm{b}$ & $\log M_\mathrm{d}$ & $\alpha_\mathrm{CO}$\\
         & ($M_\odot$) & ($M_\odot$) & [$M_\odot\,\mathrm{\left(K\,km\,s^{-1}\,pc^2\right)^{-1}}$] & & ($M_\odot$) & ($M_\odot$) & [$M_\odot\,\mathrm{\left(K\,km\,s^{-1}\,pc^2\right)^{-1}}$]\\
        \hline
        A.1 & (10.5,\,11.5) & (10.1,\,11.1) & (0,\,20) & & $10.71_{-0.08}^{+0.07}$ & $10.60_{-0.33}^{+0.33}$ & $1.50_{-0.47}^{+0.44}$\\
        A.2 & (9.5,\,12.5) & (9.1,\,12.1) & (0,\,20) & & $10.70_{-0.10}^{+0.08}$ & $10.46_{-0.84}^{+0.60}$ & $1.55_{-0.49}^{+0.47}$\\
        A.3 & 10.96 & 10.64 & (0,\,20) & & 10.96 & 10.64 & $0.04_{-0.03}^{+0.06}$\\
        B.1 & (8,\,15) & (8,\,15) & $4.3\pm1.0$ & & $10.53_{-0.24}^{+0.14}$ & $11.05_{-1.37}^{+0.38}$ & $2.28_{-0.51}^{+0.49}$\\
        B.2 & (8,\,15) & (8,\,15) & 4.34 & & $9.73_{-0.56}^{+0.24}$ & $11.63_{-0.14}^{+0.11}$ & 4.34\\
        C & (8,\,15) & (8,\,15) & (0,\,20) & & $10.73_{-0.09}^{+0.07}$ & $9.97_{-1.23}^{+0.90}$ & $1.44_{-0.46}^{+0.44}$\\
        D.1 & (10.5,\,11.5) & (10.1,\,11.1) & (0,\,20) & & $10.72_{-0.08}^{+0.07}$ & $10.58_{-0.33}^{+0.34}$ & $1.21_{-0.43}^{+0.40}$\\
        D.2 & (9.5,\,12.5) & (9.1,\,12.1) & (0,\,20) & & $10.74_{-0.07}^{+0.06}$ & $10.17_{-0.71}^{+0.67}$ & $1.12_{-0.43}^{+0.39}$\\
        D.3 & (8,\,15) & (8,\,15) & (0,\,20) & & $10.74_{-0.08}^{+0.06}$ & $9.88_{-1.20}^{+0.95}$ & $1.13_{-0.44}^{+0.43}$\\
        \hline
    \end{tabular}
    \justify{\textsc{Note} --- Prior constraints and posterior results of fitting. The uniform prior limits of parameters are denoted as `(lower, upper)'. The Gaussian priors of parameters are denoted as $\mathrm{\mu\pm\sigma}$. The fixed prior parameters are denoted as an individual number. Other parameters, $r_\mathrm{e,b}$, $n$, $r_\mathrm{e,d}$, $f_*$ and $c$ have the same prior distribution with their values adjusted in section \ref{4.3}.}
    \label{tableB1}
\end{table*}

\begin{figure}
    \centering
    \includegraphics[width=\linewidth]{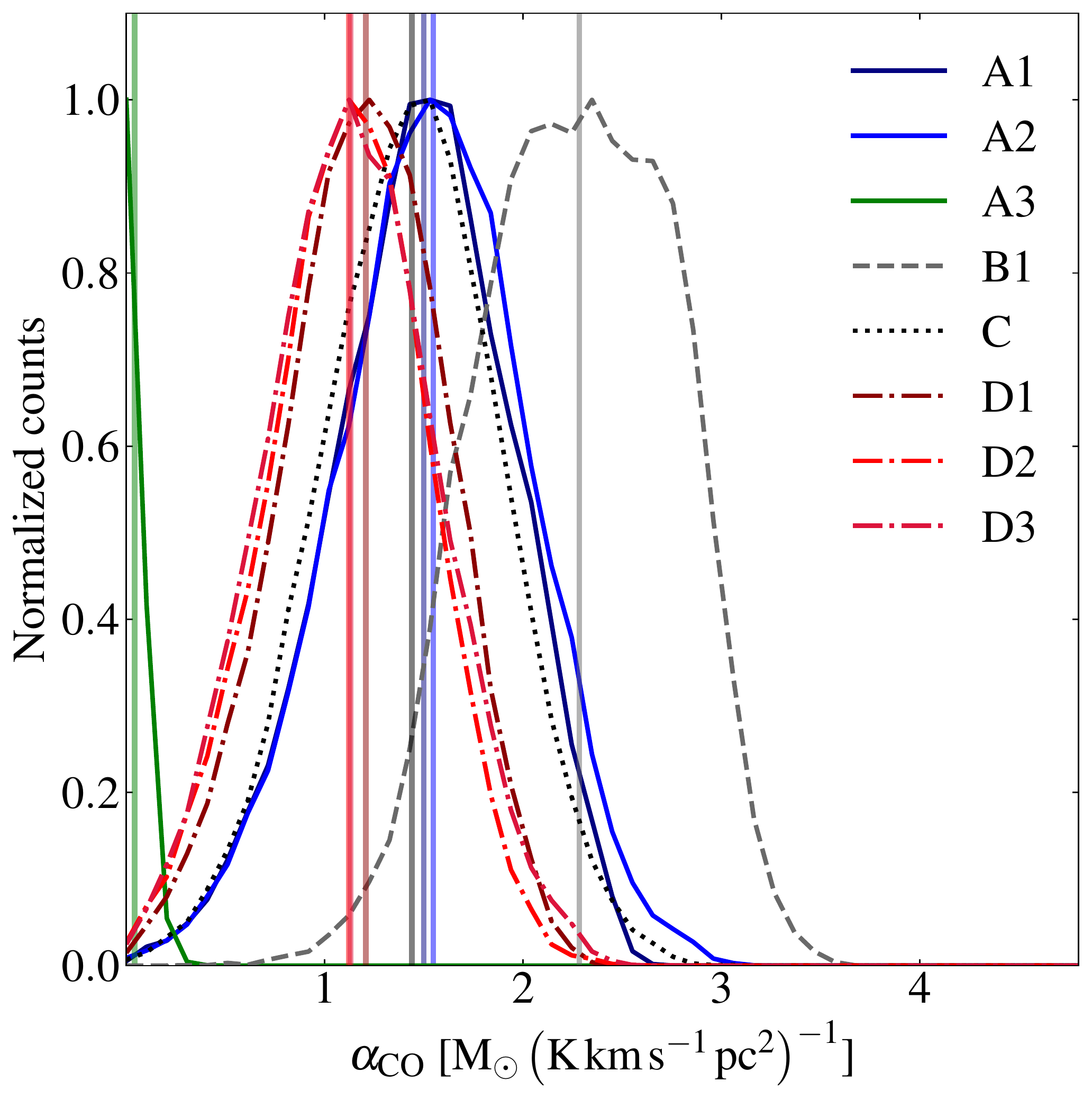}
    \caption{The posterior distribution function of \aco\ of eight cases, which have different prior constraints. The vertical lines represent the median value of \aco\ distribution, which are also shown in Table \ref{tableB1}.}
\end{figure}

In this section, We try to fit the mass of the stellar bulge, stellar disk, and the CO-to-H$_2$ conversion factor by optimizing different prior constraints. In order to the degeneracy of different components and different initial guess of parameters, we consider four main situations with total 9 cases:

\begin{enumerate}
    \item [A.] We set \aco\ as a free parameter in the fitting and limit the stellar mass within the lower and upper limits of the stellar mass estimate \cite{2021ApJ...911...94Z}.
    \item [B.] We set stellar mass as a free parameter and constrain the fitting range of \aco.
    \item [C.] We try to fit $M_b$, $M_d$, and \aco\ simultaneously with larger parameter spaces, thus those parameters are free.
    \item [D.] We fit $M_b$, $M_d$, and \aco\ without applying the asymmetric drift correction, to evaluate how significant pressure gradient support against self-gravity is in this object.
\end{enumerate}

In each case, $r_\mathrm{e,b}$, $n$, $r_\mathrm{e,d}$, $f_*$ and $c$ share the similar prior constraints (see Table \ref{table3}). In case A.1, we assume Gaussian priors for $\log (M_b/M_\odot)$ with a centered value adopted from \cite{2021ApJ...911...94Z} and a standard deviation of 0.5. In case A.2, the adopted Gaussian prior is similar to that in case A.2, while the standard deviation is three times larger. In case A.3, we fix the mass of each stellar component and study the \aco\ value. In case B.1, we assume a Gaussian prior for \aco\ with a standard deviation of 1 that is centered on the MW-like \aco\ value, and we bound each stellar component mass within $\log (M_*/M_\odot)\in[8,15]$. In case B.2, we fix the \aco\ value to that of the MW and fit the stellar mass. In the case of C, we only bound the stellar mass and let \aco\, without any further prior assumption, e.g., Gaussian distribution. In the case of D, we bound the stellar mass and \aco, but fit the rotation velocities without the asymmetric drift correction. All nine case conditions and their fitting results are listed in Table. \ref{tableB1}.

We find that the \aco\ value in case D is relatively smaller than that in cases A and C by a factor of $\sim0.75$, which indicates the effect of asymmetric drift correction. We also find that in case A.3, when we fix stellar mass, \aco\ has an extremely low value that is immoderate. And in case B, if we adopt a MW-like value, the rotation velocity is dominated by molecular gas components. This case leaves very little room for the stellar bulge in the central region. The stellar bulge mass is less than 10 percent of the value derived from the stellar continuum image. This also results in a large stellar disk mass of $10^{11}$ solar mass to account for the rotation velocity in the outer part. This result requires a  mass-to-light ratio that is different from the values adopted in \cite{2021ApJ...911...94Z}, based on the B and I band color. Thus, the MW \aco\ value in case B is unlikely to be a good assumption. Case A.1 and A.2 with a much lower \aco\ value present a more reasonable fitting for both the gas and stellar masses. As a consequence, we find that the ULIRG-like value \aco\ is reasonable in this quasar host galaxy.

\bibliography{bibliography}
\bibliographystyle{aasjournal}
\end{document}